\documentclass[10pt,journal,twoside,letterpaper]{IEEEtran}
\usepackage{cite}
\usepackage[caption=false,font=footnotesize]{subfig}
\usepackage{amsmath,amssymb,amsfonts}
\usepackage{bbm}
\providecommand{\ind}{\mathbbm{1}}
\usepackage{amsthm}
\usepackage{algorithmic}
\usepackage{algorithm}
\usepackage{graphicx}
\usepackage{epstopdf}
\usepackage{textcomp}
\usepackage{xcolor}
\usepackage{booktabs}
\usepackage{booktabs}
\usepackage{enumitem}
\usepackage{amsthm}
\newtheorem{prop}{Proposition}

\newcommand{\Ihat}{\hat{I}}

\begin{document}
\bstctlcite{IEEEexample:BSTcontrol}

\title{Drift-Resilient Predictive Interference Management: Online Adaptation of an Extreme-Value Predictor}

\author{
Aamir~Mahmood,~\IEEEmembership{Senior Member,~IEEE,}
Fateme~Salehi,~\IEEEmembership{Member,~IEEE,}   
Nurul Huda Mahmood,~\IEEEmembership{Member,~IEEE,}\\
Song Han,~\IEEEmembership{Member,~IEEE}
\vspace{-22pt}

\thanks{A. Mahmood is with the Department of Computer and Electrical Engineering, Mid Sweden University, Holmgatan 10, 851 70 Sundsvall, Sweden (e-mail: aamir.mahmood@miun.se).}
\thanks{F. Salehi is with the Department of Electronic Systems, Aalborg University, Frederikskaj 12, 2450 K\o{}benhavn SV, Denmark (e-mail: fsalehi@es.aau.dk).}
\thanks{N. H. Mahmood is with CWC, University of Oulu, Finland (e-mail: nurulhuda.mahmood@oulu.fi).}
\thanks{S. Han is with the School of Computing, University of Connecticut, Storrs, CT 06269, USA (e-mail: song.han@uconn.edu).}
\vspace{-8pt}
}

\maketitle

\begin{abstract} 
Predictive interference management supports ultra-reliable low-latency communication (URLLC) by allocating finite-blocklength resources against a high conditional quantile of next-slot
interference. Building on an extreme-value-theory/kernel-density-estimation (EVT/KDE) predictor with quantile-defined interference states, this paper makes the predictor resilient to unannounced shifts in the interference distribution. Its cumulative refitting on all past observations does not specify how the state partition and retained samples should evolve after a regime change. We show that drift first makes the quantile partition inefficient: cumulative refitting preserves tail support and coverage but cannot recover conditional resolution, and pays for it in channel use, whereas discarding history starves the tail estimator. We therefore propose event-triggered quantile-space adaptation. A nonparametric cumulative sum (CUSUM) on the predictor's level stream detects change. At each confirmed event, the quantile partition is re-derived from post-change observations, retained history is re-binned under it, and each state is refit from a bounded per-state memory, capped to balance stale-regime bias against tail-estimation variance, with a closed-form L-moment tail estimator. At a nominal coverage target $\eta\!=\!0.99$, relative to matched cumulative refitting with fixed states, the proposed method reduces channel use by $5.7$--$13.2\%$ and quantile pinball loss by $25$--$50\%$ across abrupt, incremental, compound, downward, and reversal drift, with coverage changing by $-0.5$ to $+1.0$ percentage points. At matched achieved coverage, it saves $4$--$18\%$ of channel uses, while resource use in load-only and stationary cases changes by less than $0.5\%$. With $K$ samples per state and $L$ states, predictor memory is bounded at $O(KL)$ rather than growing as $O(t)$.
\end{abstract}


\vspace{-8pt}
\section{Introduction}\label{sec:intro}
Ultra-reliable low-latency communication (URLLC) is a defining service class of fifth-generation cellular networks and beyond, with reliability targets at or below $10^{-5}$ within sub-millisecond air-interface
latency~\cite{bennis2018tail}. 
While classical wireless reliability relies on retransmissions, the sub-millisecond latency budget of URLLC often precludes even a single retransmission round, necessitating that reliability targets be met on the first attempt.

To meet a single-shot reliability target, the transmitter must provision its resource allocation, a finite blocklength of channel uses, against an upper conditional quantile of the interference in the upcoming slot. Allocating against the mean instead misses the URLLC error floor by orders of magnitude. The main idea is to replace a point estimate of the next-slot aggregate interference $I$ with a predictive upper bound $\Ihat$ at a target coverage level $\eta$, designed so that $\Pr(\Ihat\!\ge\!I)\!\ge\!\eta$, and to allocate the blocklength assuming interference at $\Ihat$. 
We define \emph{coverage} as the realized frequency of the event $\Ihat\!\ge\!I$. The metric is asymmetric: under-prediction leaves the allocation short of the reliability target, whereas over-prediction merely consumes excess resources. 

This paper examines how nonstationary interference affects the coverage contract. In~\cite{salehi2025evt}, we estimate the state space and tail model from a training trace assumed to be representative of deployment conditions. Although the formulation allows periodic parameter updates, the quantile-based state space underlying all conditional estimates remains fixed.
Yet real radio environments rarely remain stationary: machines turn on and off, devices move between cells, and traffic loads vary across shifts and days, often without explicit notification to the link layer.
A predictor that updates its parameters but not its state representation keeps conditioning on bins derived from an obsolete distribution, and the transmitter allocates against a conditional quantile of reduced resolution. We quantify how drift degrades coverage, separate the loss due to a stale state space from that due to stale samples, and develop a bounded-memory adaptation that re-derives the state space at detected changes while retaining, rather than discarding, the observations that tail
estimation requires.

\vspace{-8pt}
\subsection{Related Work}
Interference prediction has been widely studied for proactive link adaptation: in~\cite{malik2020nbiot}, neighboring links exchange scheduling information so that each can anticipate the interference the others will create; in~\cite{schmidt2021interference}, a Kalman filter with offline-parameterized gains tracks the slowly
varying interference power; and in~\cite{kim2019sense}, the transmitter senses the channel to infer receiver-side interference. These methods provide point estimates, adequate for throughput-oriented adaptation but not capturing the tail statistics governing URLLC reliability.
 
Risk-sensitive prediction for URLLC instead targets a high conditional quantile. Kernel-density-based link adaptation is proposed in~\cite{brighente2022interference}, while discrete-time Markov chain (DTMC) predictors~\cite{mahmood2020predictive,salehi2023reliable} estimate the quantile from the empirical transition structure. In~\cite{salehi2025evt} we use an EVT/KDE mixture with a KDE bulk and a generalized Pareto tail, extending prediction into extreme quantiles beyond the empirical mass. The EVT-embedded quantile transformer in~\cite{gautam2025evt} learns interference-tail dynamics in 6G industrial sub-networks, with calibrated coverage guarantees. These works either assume that the training distribution remains
representative in deployment or, as in~\cite{salehi2025evt}, periodically refit the model on all observations to date.
However, these risk-sensitive predictors do not address how to detect distributional changes during operation, which observations to use for refitting, or whether the quantile-based state space itself needs to be reconstructed.
 
The closest prior work is the nonstationary EVT framework of~\cite{mehrnia2024nonstat}, which analyzes a single received-power process offline, testing nonstationarity over a stored trace and partitioning it into piecewise-stationary groups keyed to an observed categorical covariate. Our setting instead concerns causal online prediction of aggregate interference from multiple latent interferers, with no observed regime label and a retraining horizon bounded by the drift timescale. 
 
The concept-drift literature offers several detector families. Parametric CUSUM \cite{page1954cusum,basseville1993detection} accumulates the log-likelihood ratio between two hypotheses and is Lorden-optimal when both are specified \cite{lorden1971procedures}, but here the post-change distribution is unknown a priori. Adaptive windowing (ADWIN) \cite{bifet2007adwin} tests whether any split of a variable-length window shows a significant mean shift, but the standard form tests a scalar mean and does not exploit the state-transition structure already maintained by the predictor. The Drift Detection Method (DDM) \cite{gama2004ddm} and its variants require a supervised error stream; packet errors provide a delayed and extremely sparse signal at nominal URLLC operation, and relying on them would require incurring the failures the system is intended to prevent. Exact Bayesian online change-point detection (BOCPD) \cite{adams2007bocpd} requires a specified hazard model and a tractable posterior-predictive update; applying it to the per-state KDE/generalized Pareto distribution (GPD) mixture would require additional approximation and calibration. The nonparametric change-detection method (NP-CDM) of~\cite{roveri2019dtmc}, a CUSUM on per-block likelihood ratios for homogeneous DTMCs, is the closest substrate: it monitors a finite-state Markov stream, calibrates against a reference chain estimated from short training data, and supplies a last-zero change
boundary. However, it treats the state space as fixed and does not model the state-conditional interference tail, whereas our quantile-defined states must follow the drifting distribution. This paper adapts NP-CDM to the predictor's own level stream and relaxes its fixed-state-space assumption in a controlled way, re-deriving the state space at the boundaries the detector supplies.

\vspace{-10pt}
\subsection{Contributions}
This paper introduces an online drift-adaptation mechanism for risk-sensitive interference prediction. It detects drift on the predictor's own state stream, re-derives the quantile state space at each confirmed change, re-bins the retained history under the new partition, and refits each state from a bounded per-state memory.
The detector and the state-space update apply to any state-conditional predictor, including the DTMC schemes \cite{mahmood2020predictive,salehi2023reliable}; the L-moment tail refit applies to any predictor with a generalized Pareto tail. The key contributions are:
\begin{itemize}\setlength\itemsep{0pt}
\item 
We show that drift makes the frozen quantile partition inefficient before it makes historical observations obsolete. The stale partition collapses post-drift mass into a few states, a resolution loss that no sample size repairs, and a predictor that neither refits nor re-derives loses coverage.
Re-deriving the partition alone, on common realizations with the same estimator, lowers pinball loss by $23$--$47\%$ and channel use by $6$--$14\%$ under 
representation-changing drift scenarios.
\item We propose event-triggered state-space adaptation with bounded per-state memory: a nonparametric CUSUM on the level stream detects change, the partition is re-derived at each confirmed event, and retained history is re-binned and refit from at most $K$ samples per state. Re-deriving the partition only at confirmed events, rather than continuously, keeps the level stream stationary within an epoch, as the detector requires. Against the cumulative refit~\cite{salehi2025evt}, it attains matched coverage at $4$--$18\%$ fewer channel uses on the five representation-changing scenarios, changes load-only and stationary resource use by at most $0.4\%$, and needs $O(KL)$ rather than $O(t)$ predictor memory.
\item We show that forgetting is a cost, not the mechanism. Discarding pre-change history consistently reduces coverage and increases pinball loss, and at matched achieved coverage, bounded retained memory also gives the better resource trade-off. A leading-order analysis (Proposition~\ref{prop:cap}) characterizes the bias--variance trade-off, and a disjoint pilot sweep selects the cap by worst-run coverage.
\item We refit with an L-moment tail estimator, LMRD threshold selection and a tail-adequacy diagnostic, extending the predictor to $\eta\!=\!0.999$, where the unconstrained shape of \cite{salehi2025evt} is numerically unstable. With bounded memory, the GPD tail serves over $99\%$ of prediction
slots, against $75$--$86\%$ for window-based refitting.
\end{itemize}


The rest of the paper is organized as follows. Sec.~\ref{sec:model} presents the system model, the EVT/KDE
predictor, and the drift model. Sec.~\ref{sec:failure} formalizes how drift degrades the state representation and the samples. Sec.~\ref{sec:method} develops the proposed mechanism: change detection, state-space re-derivation, bounded re-binning, and the L-moment tail refit. Sec.~\ref{sec:results} presents the Monte Carlo
evaluation, and Sec.~\ref{sec:concl} concludes.

\vspace{-8pt}
\section{System Model and Predictor Preliminaries}\label{sec:model}


\vspace{-6pt}
\subsection{Aggregate interference and traffic}\label{ssec:sysmodel}
A desired link operates in the presence of $N$ independent interferers. Time is discretized into mini-slots $t$ of fixed duration. With all received powers normalized by the receiver noise power, the instantaneous aggregate interference at the desired receiver is the interference-to-noise ratio (INR)
\begin{equation}
I_{t}=\sum_{k=1}^{N} p_{k,t}\,|h_{k,t}|^{2}\,\delta_{k,t},
\label{eq:I}
\end{equation}
where $p_{k,t}$ is the mean INR of interferer $k$, $h_{k,t}$ is its unit-power Rayleigh fading gain, and $\delta_{k,t}\!\in\!\{0,1\}$ is an activity indicator governed by a Bernoulli activation
process: each interferer starts a new message in any idle slot with probability $\mu\!\in\![0,1]$, the activity factor, and holds it for $\zeta$ slots, so $\delta_{k,t}$ is not independent across slots. An interferer may itself be a link or a network, and $\delta_{k,t}$ captures its medium-access behavior through $\mu$ and $\zeta$. The gain $h_{k,t}$ is low-pass filtered over $f_c$
slots for temporal correlation. Hence $I_t$ is auto-correlated, and drift can enter through $N$, $p_k$, $\mu$ or the fading distribution; the scenarios in Sec.~\ref{sec:results} vary the first three.

\subsection{Quantile state space and conditional interference
spaces}
\label{ssec:dtmc}
\begin{figure*}[!t]
\centering
\includegraphics[width=0.85\textwidth]{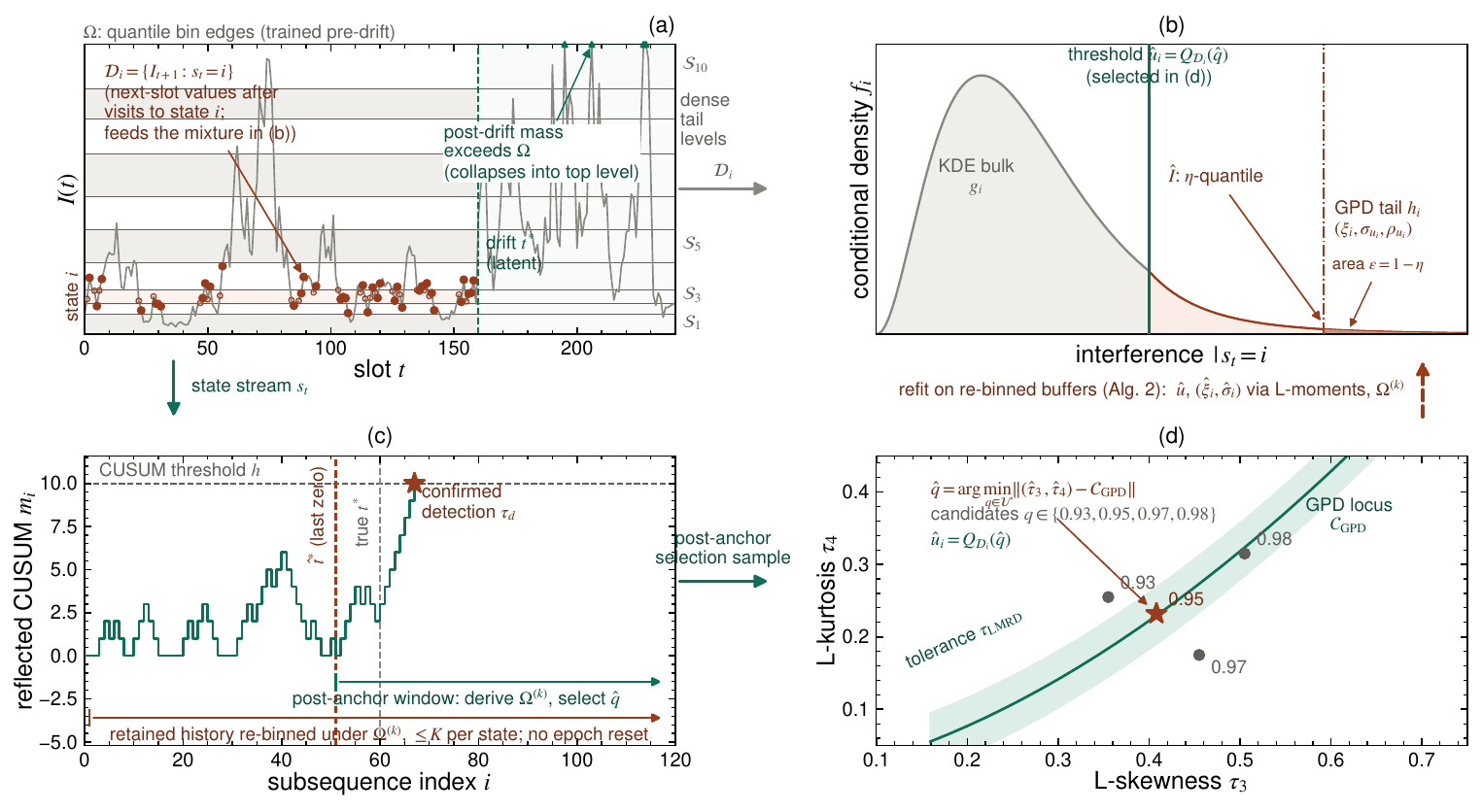}
\vspace{-12pt}
\caption{Statistical pipeline. (a)~Quantile state space: visits $s_{t}\!=\!i$ (open) and next-slot values $\mathcal{D}_{i}$ (filled); after an upward drift onset $t^{\ast}$, mass overflows the frozen $\Omega$ into the top bin. (b)~Per-state mixture, KDE bulk and GPD tail, with $\Ihat$ the $\eta$-quantile \eqref{eq:evt}. (c)~NP-CDM CUSUM on the state stream. A confirmed detection triggers the epoch update, in which $\Omega^{(k)}$ and $\hat q$ are derived from the post-anchor window and retained history is re-binned under $\Omega^{(k)}$ and kept to $K$ samples per state. (d)~LMRD selection of $\hat q$ on the L-moment-ratio plane $(\hat\tau_{3},\hat\tau_{4})$  \eqref{eq:lmrd_u}; the tolerance band is the validation diagnostic \eqref{eq:lmrd_validate}.}
\label{fig:pipeline}
\vspace{-16pt}
\end{figure*}

Risk-sensitive prediction discretizes the interference range into $L$ levels at empirical quantiles of the training trace $\mathcal{T}$. For design probabilities $0\!=\!q_{0}\!<\!q_{1}\!<\!\cdots\!<\!q_{L}\!=\!1$, the level edges are $\omega_{j}\!=\!Q_{\mathcal{T}}(q_{j})$, where $Q_{A}(q)$ is the empirical $q$-quantile of a sample set $A$, and $\Omega\!=\!(\omega_{0},\dots,\omega_{L})$ collects them, with level $i$ covering $[\omega_{i-1},\omega_{i})$. The design probabilities are narrower near $1$, so most levels
resolve the upper tail, where reliability risk concentrates; as the
interference density is low there, these levels span wider
power intervals (Fig.~\ref{fig:pipeline}a). The current level 
\begin{equation}
s_t=\ell(I_t;\Omega)=\sum_{j=1}^{L} j\,
\ind\{\omega_{j-1}\!\le\!I_t\!<\!\omega_j\}\in\{1,\dots,L\}
\label{eq:statemap}
\end{equation}
is the state on which prediction conditions, with the boundary conventions $\ell(I;\Omega)\!=\!L$ for $I\!\ge\!\omega_{L}$ and $\ell(I;\Omega)\!=\!1$ for $I\!<\!\omega_{0}$. Two design choices matter later: $\Omega$ is chosen once, as a property of an operating distribution assumed stable, and the next-slot distribution is estimated separately within each level.

Prediction conditions on the current level through the conditional samples $\mathcal{D}_{i}=\{I_{t+1}:\ s_{t}=i\}$, $i=1,\dots,L$; a first-order DTMC with transition matrix $P$, $p_{ij}=\Pr(s_{t+1}\!=\!j\mid s_{t}\!=\!i)$, summarizes this structure, its row $(p_{i1},\dots,p_{iL})$ being the discretized conditional distribution of $\mathcal{D}_{i}$.

For a target confidence $\eta\!\in\![0,1)$, risk-sensitive prediction requires the predicted interference to bound the realization from above, conditionally on the current state~\cite{mahmood2020predictive},
\begin{equation}
\Pr\big(I_{t+1} \le \Ihat \,\big|\, s_{t}=i\big) \;\ge\; \eta,
\label{eq:risk}
\end{equation}
where $\Ihat$, the predicted interference for slot $t+1$, estimates the conditional $\eta$-quantile from $\mathcal{D}_i$. Its long-run marginal rate $\bar c\!=\!\Pr(\Ihat\!\ge\!I)$ is the coverage, and the design goal is $\bar c\!\ge\!\eta$. The DTMC schemes solve
\eqref{eq:risk} by reading the quantile off the cumulative transition row, $\Ihat=\omega_{k^{\ast}}$ with
$k^{\ast}=\min\{k:\sum_{j\le k}p_{ij}\ge\eta\}$, so the prediction is always a bin edge of $\Omega$, capped at the largest interference observed in training.

\subsection{The EVT/KDE mixture predictor}
\label{ssec:mixture}
\subsubsection{Mixture model and extreme quantile}
The mixture predictor instead
fits each conditional space $\mathcal{D}_{i}$ with a semi-parametric density that extrapolates beyond the training maximum through a parametric tail. It combines a nonparametric bulk below a level-specific exceedance
threshold $u_{i}$ with an extreme-value tail above it (Fig.~\ref{fig:pipeline}b),
\begin{equation}
f_{i}(x)\;=\;
\begin{cases}
(1\!-\!\rho_{u_i})\,g_{i}(x), & x<u_{i},\\[1pt]
\rho_{u_i}\,h_{i}(x-u_{i}), & x\ge u_{i},
\end{cases}
\label{eq:mixture}
\end{equation}
with $u_{i}$ at a fixed percentile of $\mathcal{D}_{i}$, $\rho_{u_i}\!=\!\Pr(I_{t+1}\!\ge\!u_i\mid s_{t}\!=\!i)$ the exceedance rate, $g_{i}$ the bulk density normalized below $u_{i}$, and $h_{i}$ the normalized exceedance density, so that $\int_{0}^{\infty} f_{i}(x)\,\mathrm{d}x=1$.

The tail branch models the conditional exceedance distribution $F_{u_i}(y):=\Pr(I_{t+1}-u_{i}\le y \mid I_{t+1}\!>\!u_{i}, s_{t}=i)$, $y\ge0$, as generalized Pareto, $F_{u_i}=G_{\xi_{i},\sigma_{u_i}}$,
\begin{equation}
G_{\xi,\sigma}(y) \;=\; 1 - \!\Big(1+\tfrac{\xi y}{\sigma}\Big)_{+}^{-1/\xi},
\quad\sigma\!>\!0,\;\xi\!\in\!\mathbb{R},
\label{eq:gpd_cdf}
\end{equation}
with level-specific shape $\xi_{i}$ and scale $\sigma_{u_i}$, the exponential limit $G_{0,\sigma}(y)=1-e^{-y/\sigma}$ at $\xi\!=\!0$, and the induced conditional tail
\begin{equation}
\Pr\big(I_{t+1}>x \,\big|\, s_{t}=i\big)=\rho_{u_i}
\Big(1+\xi_{i}\,\tfrac{x-u_{i}}{\sigma_{u_i}}\Big)_{+}^{-1/\xi_{i}}\!\!,
\!\! x\ge u_{i}.
\label{eq:tail_surv}
\end{equation}
The GPD is the peaks-over-threshold (PoT) limit of EVT~\cite{coles2001evt}.
The Rayleigh aggregate \eqref{eq:I} is asymptotically Gumbel ($\xi\!=\!0$), and at finite thresholds the fitted shapes sit just above it, $\hat\xi\!\in\![0,0.5]$~\cite{salehi2025evt}. Below $u_{i}$, the bulk is a Gaussian KDE on the $N_{\mathrm b}$ sub-threshold samples of $\mathcal{D}_{i}$,
\begin{equation}
\tilde g_{i}(x)\;=\;\frac{1}{N_{\mathrm b}}\sum_{n=1}^{N_{\mathrm b}}
\frac{1}{\sqrt{2\pi}\,T}\;e^{-(x-x_{n})^{2}/(2T^{2})},
\label{eq:kde}
\end{equation}
with bandwidth $T$ minimizing the asymptotic mean integrated squared error~\cite{botev2010kde}; $g_{i}$ is its renormalized truncation below $u_{i}$.

Inverting \eqref{eq:tail_surv} at survival level $1\!-\!\eta$ gives the standard GPD extreme quantile~\cite{coles2001evt}
\begin{equation}
\Ihat \;=\; u_i+\frac{\sigma_{u_i}}{\xi_i}\!\left[\Big(\frac{\rho_{u_i}}
{1-\eta}\Big)^{\!\xi_i}\!-1\right],
\label{eq:evt}
\end{equation}
with limit $\Ihat=u_i+\sigma_{u_i}\ln(\rho_{u_i}/(1\!-\!\eta))$ at $\xi_i\!=\!0$. When the requested quantile lies below $u_{i}$, $\Ihat$ is the corresponding quantile of the KDE-smoothed conditional distribution. The GPD branch is active whenever $\eta\!>\!1\!-\!\rho_{u_i}$, i.e., whenever the target exceeds the threshold percentile.

\subsubsection{Threshold choice and sample scarcity}
In~\cite{salehi2025evt}, $u_{i}$ is the $97$th percentile of $\mathcal{D}_{i}$. The percentile above which the GPD holds is distribution-dependent and cannot follow a regime change when fixed, so Sec.~\ref{ssec:lmrd} selects it from the data at each epoch update. Irrespective of the threshold, the tail fit is starved by
construction. With $n_{0}$ the training-window length and $\pi_{i}$ the visit probability of state $i$, the expected exceedance count above $u_{i}$ is $\mathbb{E}[N_{u,i}]\!\approx\!n_{0}\pi_{i}\rho_{u_i}$, about two
per state at $n_{0}\!=\!10^{3}$, $L\!=\!15$ and $\rho_{u_i}\!\approx\!0.03$, and fewer in the rarely visited
high-interference states. Since the largest of $n$ observations estimates at most the $(1\!-\!1/n)$-quantile, an empirical $\eta\!=\!0.99$ quantile needs $100$ observations, which many conditional spaces lack. The requested quantile is then extrapolated through the fitted tail, with standard error decreasing as $N_{u,i}^{-1/2}$ in the exceedance count $N_{u,i}$. Tail accuracy is thus governed by how many exceedances each state accumulates.

\subsection{Finite-blocklength resource allocation}\label{ssec:cov}
With $\bar\gamma$ the mean SNR, known at the transmitter, and $\Ihat$ the predicted interference, the transmitter operates against the predicted SINR $\hat\gamma=\bar\gamma/(1\!+\!\Ihat)$. At URLLC blocklengths, the blocklength $M$, the number of channel uses required to deliver $b$ information bits at the decoding-error
target $\epsilon$, is~\cite{polyanskiy2010fbl}
\begin{equation}
M\!\approx\!\frac{b}{C(\hat\gamma)}\!+\!\frac{Q^{-1}(\epsilon)^{2}V(\hat\gamma)}
{2\,C(\hat\gamma)^{2}}\!\left[1\!+\!\sqrt{1\!+\!\frac{4b\,C(\hat\gamma)}
{Q^{-1}(\epsilon)^{2}V(\hat\gamma)}}\right]\!,
\label{eq:M}
\end{equation}
with capacity $C(\gamma)=\log_{2}(1\!+\!\gamma)$, dispersion $V(\gamma)=(\log_{2}\!e)^{2}[1\!-\!(1\!+\!\gamma)^{-2}]$, and $Q^{-1}$ the inverse Gaussian Q-function. When $\Ihat\!<\!I_{t+1}$, the $M$ selected for $\hat\gamma$ no longer certifies $\epsilon$; when $\Ihat\!\ge\!I_{t+1}$, the target is met at a spectrum cost. If the allocation certifies $\epsilon$ whenever $\{I_{t+1}\!\le\!\Ihat\}$ holds, the unconditional block-decoding error probability satisfies $\Pr(\mathcal{E}_{t+1})\!\le\! 1-\eta(1-\epsilon)\!\le\! (1-\eta)+\epsilon$. A deployment with total error target $\epsilon_{\mathrm{tot}}$ must therefore split its budget, $(1-\eta)+\epsilon\le     \epsilon_{\mathrm{tot}}$. For a given $\eta$, an end-to-end $10^{-5}$ certificate would require $\eta\to1-O(\epsilon)$; the values of $\eta$ used for comparing adaptation methods are those the sample sizes resolve statistically (Sec.~\ref{sec:results}).

\vspace{-10pt}
\subsection{Concept drift and its detection}\label{sec:cd}
\label{sec:drift}
Let $\mathcal{P}_{t}(s,I_{t+1})\!=\!\Pr_t(s_t)\Pr_t(I_{t+1}\!\mid\! s_t)$ denote the joint distribution of the current state $s_{t}$ and the next-slot interference $I_{t+1}$, that is, how often each state is visited together with the distribution of $I_{t+1}$ given $s_{t}$. Concept drift~\cite{gama2014survey} is a change in $\mathcal{P}_{t}$ during the lifetime of the model, beginning at an unknown instant $t^{\ast}$ and driven by the latent mechanisms of \eqref{eq:I}, alone
or combined, so the interference stream is the transmitter's only evidence of it.
By time profile~\cite{gama2014survey}, drift is \emph{sudden}, \emph{incremental}, or \emph{recurring}.


A change detector monitors a statistic of the live stream against a reference learned in training and raises an alarm when a cumulative discrepancy crosses a threshold~\cite{page1954cusum,basseville1993detection}. Two quantities govern any such detector: the \emph{average run length under the null} ($\mathrm{ARL}_{0}$), the expected time between false alarms without drift, and the \emph{detection delay} between a true change and its alarm. For a parametric CUSUM with known log-likelihood ratio at threshold $h$, Lorden's bound~\cite{lorden1971procedures} gives $\mathrm{ARL}_{0}\!\gtrsim\!e^{h}$ while the delay grows only linearly in $h$, so false alarms can be made exponentially rare at linear cost in delay.

\vspace{-6pt}
\section{Silent Loss of Coverage Under Concept Drift}\label{sec:failure}

\subsection{Stale quantities after drift}
A predictor trained on the pre-drift trace evaluates~\eqref{eq:evt} from four stale quantities. (i) The state map $\ell(\cdot;\Omega)$ keeps the pre-drift edges, so post-drift mass piles into a few levels and one state label now covers widely different $I_{t}$, a resolution loss no sample size repairs. (ii) The conditional KDE bulk no longer matches the new bulk percentiles. (iii) The GPD tail parameters $(u_i,\sigma_{u_i},\xi_i,\rho_{u_i})$ are fit to old exceedances. (iv) The transition statistics of the level sequence no longer match training. A downward shift leaves the old predictor conservative at a resource cost, while an upward or tail-heavier shift leaves $\Ihat$ below the post-change quantile.

\vspace{-7pt}
\subsection{Pre-drift coverage at the URLLC sample budget}
Even under stationarity, the EVT/KDE predictor of~\cite{salehi2025evt}, trained once on its $10^{3}$-sample budget and never refit, reaches only $\bar c\!=\!0.965$, short of the target $\eta\!=\!0.99$. This is a finite-sample effect, not drift: a $4\!\times\!10^{3}$-sample budget raises stationary coverage to $0.985$, yet the frozen predictor still falls to $0.853$ under an abrupt interferer join.

\vspace{-8pt}
\subsection{Silent collapse}
After $t^{\ast}$, a predictor that neither re-derives $\Omega$ nor refits keeps evaluating \eqref{eq:evt} with the stale quantities, and its coverage falls to $0.59$--$0.75$ under the drift scenarios of Sec.~\ref{sec:results} (Table~\ref{tab:main}), below $0.5$ when the interference scale changes.
The failure is silent. The predictor still returns a plausible quantile and the allocator a blocklength near or below that of an oracle knowing $I_{t+1}$, so nothing signals that the model is invalid. Yet the achieved outage stays orders of magnitude above the target however tightly $\epsilon$ is set (Fig.~\ref{fig:conv}). The periodic refit of~\cite{salehi2025evt} on all past observations does not collapse (coverage $0.96$--$0.99$, and $0.987$ on the stationary control). Under \emph{representation-changing} drift, which misaligns the level distribution with the pre-drift edges, cumulative refitting preserves coverage but cannot restore conditional resolution, while its memory grows as $O(t)$.
The question is not whether to refit, but what to re-derive and what to retain.

\vspace{-8pt}
\subsection{Why passive adaptation is insufficient}
Refitting on a sliding window of length $L_w$, the standard passive adaptation~\cite{gama2014survey} and the natural alternative to the cumulative refit of~\cite{salehi2025evt} do not resolve the failure mode.
\begin{itemize}[leftmargin=*]
\setlength{\itemsep}{1pt}
\item \emph{Transient contamination:} for $\sim\!L_w$ samples after $t^{\ast}$, and under incremental drift for as long as the distribution changes, the refit fits a mixture of both regimes.
\item \emph{Noisy tail at small $L_w$:} a window short enough to flush a drift leaves state $i$ only $\approx\!L_w\pi_i\rho_{u_i}$ exceedances for tail fit.
\item \emph{Stationary penalty:} without drift, the $L_w$-sample refit is poorer than the trained model.
\item \emph{Stale partition under cumulative refit:} refitting on all data removes the variance penalty but not the resolution loss, at excess channel use and $O(t)$ memory.
\end{itemize}
What is missing is an explicit event at which $\Omega$ is re-derived and retained observations are re-labeled, rather than discarded.

\section{Drift-Aware Adaptation}\label{sec:method}

\begin{figure}[t]
    \centering
    \includegraphics[width=0.95\linewidth]{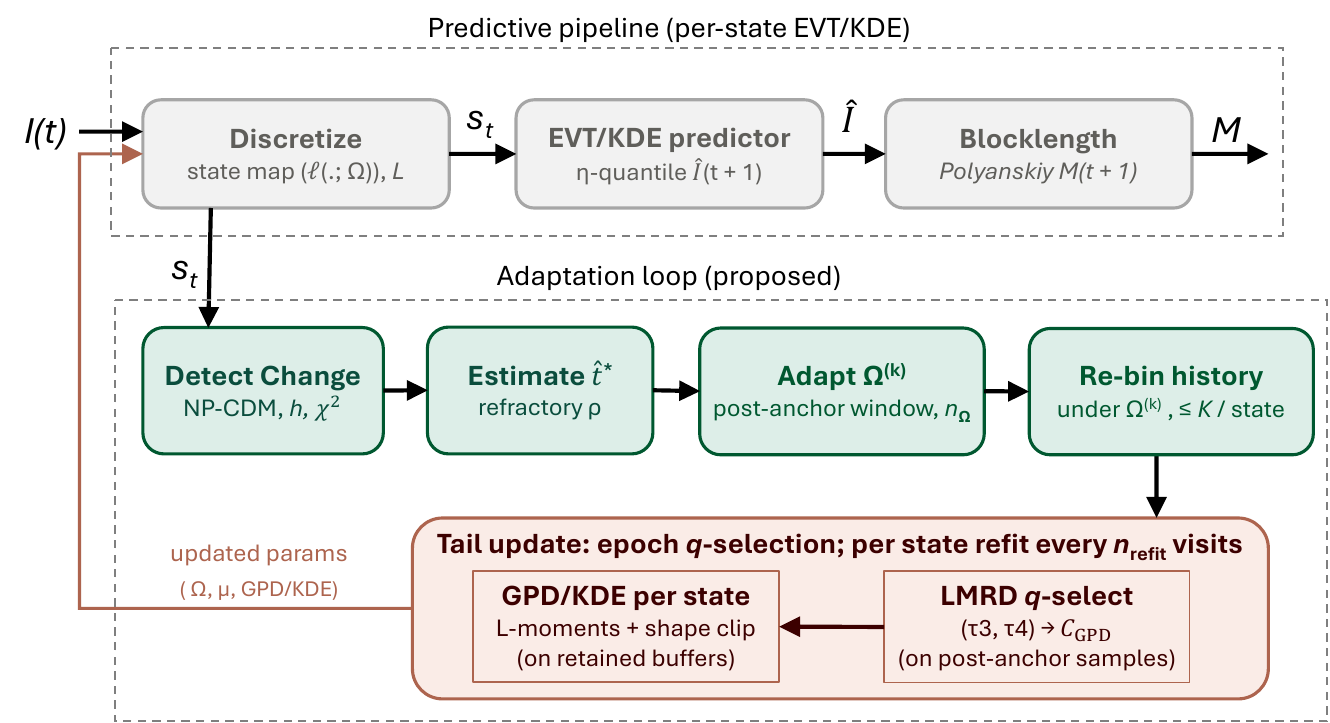}
    \vspace{-8pt}
    \caption{Proposed architecture as the control-flow
complement of Fig.~\ref{fig:pipeline}.}
    \label{fig:arch}
\vspace{-6pt}    
\end{figure}

Fig.~\ref{fig:arch} shows the proposed architecture. The upper block is the prediction pipeline, executed every slot and initialized from a stationary training window, comprising the discretization \eqref{eq:statemap}, per-state EVT/KDE mixture \eqref{eq:mixture} returning $\Ihat$ \eqref{eq:evt}, and the
blocklength \eqref{eq:M}. The lower block is the adaptation loop we add. A sequential detector monitors the predictor's level sequence $s_t$, declares a change when its dynamics depart from a reference chain learned in training, and localizes when the change began (Sec.~\ref{sec:method}-A). At each confirmed change, the state space is re-derived from post-change observations (Sec.~\ref{sec:method}-C), and the retained history is
re-binned under it and refit from at most $K$ samples per state (Sec.~\ref{sec:method}-D), with the threshold and GPD parameters re-selected by the L-moment apparatus at each refit (Sec.~\ref{ssec:lmrd}). Three constraints of the URLLC setting shape the design: i) A realized link outage, the event the predictor
exists to avoid, cannot serve as a supervised drift signal; ii) A stale $\Omega$ coarsens the conditional representation regardless of the fitting-sample size, so the representation must follow the regime; and iii) The GPD estimator is limited by its exceedance count, so pre-change observations are worth retaining under the new partition, within a memory bound to keep the predictor deployable.

\subsection{Sequential drift detection on the level stream}
\begin{figure}[!t]
\centering
\includegraphics[width=1\columnwidth]{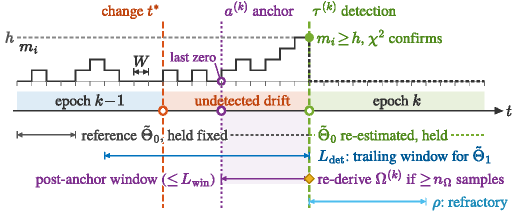}
\vspace{-20pt}
\caption{Detector and epoch-update timing: the sign CUSUM $m_i$ \eqref{eq:cusum} over blocks of $W$ slots, the reference $\tilde\Theta_{0}$, fixed within the epoch, and the trailing window $L_{\mathrm{det}}$ for $\tilde\Theta_{1}$, a confirmed detection $\tau^{(k)}$ anchored at the last zero $a^{(k)}$ \eqref{eq:cp}, the post-anchor window from which $\Omega^{(k)}$ is re-derived at detection if it holds at least $n_{\Omega}$ samples \eqref{eq:adaptomega}, and the refractory $\rho$ \eqref{eq:epoch}.}
\label{fig:timeline}
\vspace{-4pt}
\end{figure}

We treat drift as a hypothesis test on the dynamics of the level stream $s_t=\ell(I_t;\Omega)$.
Monitoring $s_t$ means the detector sees exactly what the predictor sees and needs no auxiliary state. The stream is finite-state and modeled as first-order Markov, so the reference statistics, at most $L^2$ transition probabilities in $[0,1]$, have bounded variance regardless of the interference tail heaviness. The detector responds to changes that alter state occupancy or transition dynamics, so a change confined within a single state goes undetected until it moves samples across a state boundary. Over a sample set $\mathcal{A}$, let $n_{ij}(\mathcal{A})$ count the observed $i\!\to\!j$ transitions; with Laplace smoothing $\beta\!>\!0$, we estimate the transition matrix and its stationary
distribution as 
\begin{equation}
\hat P_{ij}(\mathcal{A})=
\frac{n_{ij}(\mathcal{A})+\beta}
{\sum_{k=1}^{L}\big(n_{ik}(\mathcal{A})+\beta\big)},\quad
\hat\pi=\hat\pi\hat P,\ \ \hat\pi\mathbf{1}=1,
\label{eq:mle}
\end{equation}
forming the estimated chain $\tilde\Theta=(\hat P,\hat\pi)$. The smoothing keeps every entry of $\hat P$ positive, so the block likelihood below is finite and $\hat\pi$ is unique. We estimate the chain twice: the \emph{reference} $\tilde\Theta_0$ is learned on the training window and the \emph{current} $\tilde\Theta_1$ is re-learned continuously on the trailing $L_{\mathrm{det}}$ slots, the detector's window (Fig.~\ref{fig:timeline}), distinct from the passive baseline's refit window $L_w$. The reference is held
fixed within an epoch and re-estimated after each accepted change from the post-anchor sample under the new partition (Sec.~\ref{sec:method}-C).

Since the post-change distribution is unknown, we adopt the nonparametric detector
of~\cite{roveri2019dtmc}, with the current chain $\tilde\Theta_1$ as the alternative. 
We adapt it to the predictor's own state stream, with smoothed plug-in likelihoods, joint calibration
of threshold and confirmation, and Page's last-zero boundary repurposed as a retraining anchor. It rests on one assumption: \emph{within an epoch, $\Omega$ is fixed and the conditional chain is locally homogeneous}. Sec.~\ref{sec:method}-C confines changes of $\Omega$ to epoch boundaries, and incremental drift violates homogeneity for as long as the distribution changes. Since a single transition carries little evidence about an
$L\!\times\!L$ matrix, the stream is scored in non-overlapping blocks of $W$ consecutive levels
\vspace{-6pt}
\begin{equation}
\mathbf{w}_i=\big(s_{W(i-1)+1},\,\dots,\,s_{Wi}\big),
\label{eq:subseq}
\end{equation}
at a latency cost of at most $W$ slots. The log-likelihood of a subsequence under a chain $\tilde\Theta=(\tilde P,\tilde\pi)$ is its start-state term plus the sum of its transition terms,
\begin{equation}
\mathcal{L}_{\tilde\Theta}(\mathbf{w}_i)=
\log\tilde\pi\big(s_{W(i-1)+1}\big)
+\!\!\sum_{j=W(i-1)+1}^{Wi-1}\!\!\log\tilde P_{\,s_j,\,s_{j+1}}
\label{eq:loglik}
\end{equation}
and we score each block by the difference
\begin{equation}
l_i=\mathcal{L}_{\tilde\Theta_1}(\mathbf{w}_i)
-\mathcal{L}_{\tilde\Theta_0}(\mathbf{w}_i),
\label{eq:llr}
\end{equation}
which is positive when the current chain explains the block better than the reference. Under stationarity, the two chains explain a block about equally well, up to a small positive bias in $l_i$ from the trailing window containing the block it scores, which the calibration below absorbs. After drift, sustained positive scores accumulate evidence against the reference.

We aggregate only the \emph{sign} of $l_i$, which bounds the influence of heavy-tailed score magnitudes without assuming a parametric
score distribution. The detection statistic is a reflected counting CUSUM~\cite{page1954cusum},
\begin{equation}
m_0=0,\qquad
m_i=\max\!\big(0,\;m_{i-1}+\operatorname{sign} l_i\big),
\label{eq:cusum}
\end{equation}
and a change is declared at $i^{\ast}=\min\{i:\,m_i\!\ge\!h\}$, the first block at which $m_i$ reaches a threshold $h$ counted in blocks; larger $h$ makes false alarms rarer at the cost of detection delay. Here, $m_i$ counts the current run of net positive evidence, the $\max(0,\cdot)$ reflects it at zero so that pre-change accumulation cannot delay a later alarm, and the reflection supplies the last-zero boundary of \eqref{eq:cp}(Fig.~\ref{fig:pipeline}c). Since parametric CUSUM bounds do not transfer to the sign-transformed statistic, $h$ is set a priori, with its sensitivity reported in Sec.~\ref{ssec:discussion}.

Since $\tilde\Theta_{0}$ is a finite-sample estimate, the CUSUM
\eqref{eq:cusum} alone over-fires on stream stretches the training window under-represents, so each candidate $i^{\ast}$ is confirmed by a two-sample chi-square ($\chi^{2}$) test on the transition counts~\cite{roveri2019dtmc}. For each level $r$, with $n_{rj}^{(0)}$ and $n_{rj}^{(1)}$ the $r\!\to\!j$ counts in the training and trailing windows, $N_{r}^{(k)}\!=\!\sum_{j}n_{rj}^{(k)}$, the pooled rate $\tilde p_{rj}=(n_{rj}^{(0)}+n_{rj}^{(1)})/(N_{r}^{(0)}+N_{r}^{(1)})$, and $\mathcal{J}_{r}=\{j:\,n^{(0)}_{rj}+n^{(1)}_{rj}>0\}$, the statistic
\begin{equation}
\chi^{2}_{r}=\sum_{k=0}^{1}\sum_{j\in\mathcal{J}_{r}}
\frac{\big(n_{rj}^{(k)}-N_{r}^{(k)}\tilde p_{rj}\big)^{2}}
{N_{r}^{(k)}\tilde p_{rj}}
\label{eq:chi2stat}
\end{equation}
is asymptotically $\chi^{2}$ with $\nu_{r}=|\mathcal{J}_{r}|-1$ degrees of freedom when both windows share one distribution. Over the tested rows $\mathcal{R}_{\mathrm{test}}=\{r:\,N^{(0)}_{r}\!>\!0,\, N^{(1)}_{r}\!>\!0,\,|\mathcal{J}_{r}|\ge 2\}$, the candidate is accepted if at least one row rejects,
\begin{equation}
\exists\,r\!\in\!\mathcal{R}_{\mathrm{test}}:\ \chi^{2}_{r}\;>\;
\chi^{2}_{1-\alpha/L,\,\nu_{r}},
\label{eq:chi2}
\end{equation}
with $\alpha/L$ the Bonferroni correction for $L$ rows. Because high-tail rows are sparse at the operating window lengths, this stage is an empirical screen, not an exact test. We report the false-alarm rate it yields with $h$ on the stationary control in Sec.~\ref{ssec:discussion}.

The retraining anchor is the slot from which post-change samples are collected for the epoch update (Sec.~\ref{sec:method}-C). It follows the last-zero rule: an earlier change would have begun a run that an intervening zero erased, so the anchor is the last block at which the CUSUM was zero, well defined since $m_{0}=0$,
\begin{equation}
\hat t^\ast \;=\; W\cdot\max\{\,j\le i^\ast:\ m_j=0\,\},
\label{eq:cp}
\end{equation}
the last slot of that block. Resolved to one block, it is a retraining boundary, not a change-point estimate; its error
is reported in Sec.~\ref{ssec:discussion}.

Raw detections $\tau_d^{(1)},\tau_d^{(2)},\dots$ are filtered by a refractory interval $\rho$, a small multiple of the typical detection delay: a confirmed detection within $\rho$ of an accepted one belongs to the same episode and resets $m_{i}\!\leftarrow\!0$ without creating a new epoch. The $k$-th accepted detection and its anchor (Fig.~\ref{fig:timeline}) are
\begin{equation}
\tau^{(k)}\!=\!\min\big\{\tau_d^{(m)}:\, \tau_d^{(m)}\!-\!\tau^{(k-1)}\!>\!\rho\big\},
\quad a^{(k)}\!=\!\hat t^{\ast(k)},
\label{eq:epoch}
\end{equation}
\vspace{-25pt}
\subsection{What drift degrades: representation versus memory}
\label{ssec:decouple}
A detection licenses two operations, and a natural design performs both at every accepted change: re-derive 
$\Omega$ from post-change observations, and discard the predictor's pre-change history. The two act on different terms of the excess risk, as the decomposition below shows, and should be decoupled.

Write $X\!=\!I_{t}$, $Y\!=\!I_{t+1}$ and $S_{\Omega}\!=\!\ell(X;\Omega)$, and let $\psi_{\eta}(e)\!=\!e\,(\eta-\mathbf{1}\{e<0\})$ be the pinball loss~\cite{koenker1978}. The expected loss $\mathcal{L}(f)\!=\!\mathbb{E}[\psi_{\eta}(Y-f)]$ is minimized by the conditional $\eta$-quantile, so under the current regime $q_{\eta}(x)\!=\!Q_{\eta}(Y\mid X\!=\!x)$ is the optimal predictor given $I_{t}$ and $q_{\eta,\Omega}(r)\!=\!Q_{\eta}(Y\mid S_{\Omega}\!=\!r)$ the optimal one given the state. With each predictor evaluated on its own input, the excess risk of the predictor's estimate $\hat q_{\eta,\Omega}$ splits as
\begin{equation}
\mathcal{L}(\hat q_{\eta,\Omega})-\mathcal{L}(q_{\eta})
=\underbrace{\mathcal{L}(q_{\eta,\Omega})-\mathcal{L}(q_{\eta})}_{\text{representation}}
+\underbrace{\mathcal{L}(\hat q_{\eta,\Omega})-\mathcal{L}(q_{\eta,\Omega})}_{\text{estimation}},
\label{eq:decomp}
\end{equation}
both terms nonnegative since $S_{\Omega}$ is a function of $X$.

The representation term is the cost of replacing $I_{t}$ by the discrete state. It depends only on the resolution of the active partition: no sample reduces it while $\Omega$ is fixed, and re-deriving $\Omega$ reduces it by redistributing the current interference mass across the states. The estimation term is the
cost of estimating the statewise quantile from retained observations, finite-sample variance, bias from earlier regimes and tail-model approximation. Proposition~\ref{prop:cap} gives its leading bias--variance behavior.

Under representation-changing drift, the pre-change edges concentrate the current distribution in a few states, broadening the range of $I_{t}$ each represents. A cumulative fixed-$\Omega$ estimator still estimates $q_{\eta,\Omega}(r)$ accurately and maintains coverage, but pays the representation loss as conservative allocation: the adaptive partition needs $6$--$13\%$ fewer channel resources (Sec.~\ref{sec:results}). Drift therefore makes the frozen quantile partition inefficient before it makes historical observations obsolete. The design principle that follows, and organizes the rest of the paper, is to adapt the state representation without discarding the tail observations on which extreme-quantile accuracy depends.

\subsection{Per-epoch adaptive state space}
After a representation-changing drift, the pre-drift edges no longer realize the prescribed quantile masses, and re-deriving $\Omega$ from the post-anchor window restores them on the post-drift distribution. At the $k$-th accepted change (Sec.~\ref{sec:method}-A), let $\mathcal{I}^{(k)}=\{I_r:\,a^{(k)}\!+\!1\!\le\!r\!\le\!\tau^{(k)}\}$ be the post-anchor samples. The per-epoch $\Omega$ applies the edge rule of Sec.~\ref{ssec:dtmc} to $\mathcal{I}^{(k)}$ in place of the training trace, with the same design probabilities $q_0,\dots,q_L$
\begin{equation}
\Omega^{(k)}=
\begin{cases}
\big[Q_{\mathcal{I}^{(k)}}(q_0),\dots,
Q_{\mathcal{I}^{(k)}}(q_{L})\big], & |\mathcal{I}^{(k)}|\ge n_\Omega,\\[2pt]
\Omega^{(k-1)}, & \text{otherwise,}
\end{cases}
\label{eq:adaptomega}
\end{equation}
where $n_{\Omega}$ is the minimum post-anchor count for re-deriving the $L$ edges. The update is attempted only at the accepted detection. If the post-anchor window holds fewer than $n_{\Omega}$ samples, the epoch still opens (steps~2--5) under $\Omega^{(k-1)}$, which holds until the next accepted event.

The $k$-th accepted change triggers the epoch update:
\begin{enumerate}
\item derive $\Omega^{(k)}$ from the post-anchor window $\mathcal{I}^{(k)}$
      by \eqref{eq:adaptomega};
\item re-estimate the detector reference chain $\tilde\Theta_0$ from the
      $\Omega^{(k)}$-discretization of the same window;
\item re-bin the retained observations under $\Omega^{(k)}$: each
      stored pair $(I_t,I_{t+1})$ joins state
      $\ell(I_t;\Omega^{(k)})$;
\item reconstruct each $\mathcal{D}_i$ from
      the re-binned samples, retaining at most
      $K$ samples per state (Sec.~\ref{ssec:memory});
\item refit the KDE/GPD mixture on the reconstructed samples with the
      LMRD rule, sample gates, shape clip and fallback (Sec.~\ref{ssec:lmrd}).
\end{enumerate}
Re-binning does not remove pre-change samples; it reassigns them to the new states, where they add tail samples but may carry stale bias. The memory cap bounds their share (Sec.~\ref{ssec:memory}). Between changes, each new observation $I_t$ is appended to $\mathcal{D}_{s_{t-1}}$, and that state's mixture is refit at the held $\hat q$ every $n_{\mathrm{refit}}$ visits.

Restricting the $\Omega$ update to epoch boundaries relaxes the fixed-state-space assumption of~\cite{roveri2019dtmc}. Updating $\Omega$ continuously would impose non-stationarity on the monitored stream that originates in the discretization rather than in the interference process. Within an epoch, $\Omega$ is therefore fixed, and under an approximately stationary regime the detector's assumptions hold. The relaxation is confined to the boundaries, where the reference reset of step~2 coincides with the operation any
exogenous change of $\Omega$ would require.

\subsection{Bounded per-state memory}\label{ssec:memory}
Given the anchor $a^{(k)}$ of Sec.~\ref{sec:method}-A, three memory policies are possible for the per-state samples. The \emph{anchored window} $[a^{(k)}\!+\!1,\,t]$, of fixed maximum length $L_{\mathrm{win}}$, re-derives $\Omega$ at each detected change but discards the pre-change history. It removes contamination but starts every epoch from a small sample, a trade-off studied in Appendix~\ref{sec:S-crossover}.
\emph{Unbounded retention} keeps every sample, as cumulative refitting does, but re-binned under each new partition, at $O(t)$ memory and with contamination decaying only as $1/t$. The \emph{proposed policy} retains the $K$ most recent samples of each state under the re-binned partition.

Fix an epoch $k$ and a state $r$ of the active partition $\Omega^{(k)}$. Immediately after the epoch update, state $r$ inherits $b_r^{(k)}\le K$ re-binned samples, of which $s_r^{(k)}\le b_r^{(k)}$ are stale, i.e.\ drawn before the current regime; let $v_r^{(k)}(t)$ be the number of visits to $r$ since the update, each appending one sample under FIFO retention at $K$. The buffer then holds $m_r^{(k)}(t)=\min\{K,\,b_r^{(k)}+v_r^{(k)}(t)\}$ samples. The buffer thus mixes two regimes, and we index their GPD parameters by $0$ (pre-change) and $1$ (post-change). The state's $\eta$-quantile shifts by $\Delta Q_{\eta,r}\!=\!Q_{\eta,1}(r)-Q_{\eta,0}(r)$, and
$\kappa_{\eta}(\xi,\rho_u)$ \eqref{eq:varQ} is the dimensionless variance factor of its estimate at target $\eta$ and exceedance rate $\rho_u$.  Proposition~\ref{prop:cap} bounds the stale fraction $\alpha_r^{(k)}(t)$, the share of the buffer still drawn from the
pre-change regime, and gives the resulting bias--variance trade in $K$.

\begin{prop}[Bounded memory: bias--variance trade]
\label{prop:cap}
Assume the probability-weighted-moment (PWM) regularity conditions of~\cite{hosking1987pwm} ($\xi<1/2$), and approximate the effect of the mixed sample to first order in the stale fraction (linear contamination). The stale fraction of the state-$r$ buffer satisfies
\begin{equation}
\alpha_r^{(k)}(t)\;\le\;
\frac{\min\big\{s_r^{(k)},\,[\,K-v_r^{(k)}(t)\,]_{+}\big\}}
     {\min\{K,\,b_r^{(k)}+v_r^{(k)}(t)\}},
\quad [x]_{+}=\max\{x,0\},
\label{eq:alpha_cap}
\end{equation}
so that $\alpha_r^{(k)}(t)=0$ as soon as $v_r^{(k)}(t)\ge K$. The
first-order MSE of the predicted quantile $\Ihat$ for state $r$ is
\begin{equation}
\mathrm{MSE}_r(t)\;\approx\;
\alpha_r^{(k)}(t)^{2}\,|\Delta Q_{\eta,r}|^{2}
\;+\;
\frac{\kappa_{\eta}(\xi_1,\rho_{u_r})\,\sigma_{1}^{2}}
     {\rho_{u_r}\,m_r^{(k)}(t)} .
\label{eq:mse_cap}
\end{equation}
For unbounded retention ($K=\infty$) the stale fraction is instead
$s_r^{(k)}/\big(b_r^{(k)}+v_r^{(k)}(t)\big)$, which decays only as
$1/v_r^{(k)}(t)$ and never vanishes in finite time.
\end{prop}
 
\begin{IEEEproof}
See Appendix~\ref{sec:analysis}.
\end{IEEEproof}
 
Eq.~\eqref{eq:mse_cap} is a leading-order, statewise quadratic approximation for the estimation term of \eqref{eq:decomp}. For state $r$, if $f_{Y\mid S_{\Omega}=r}$ is positive and continuous at $q_{\eta,\Omega}(r)$, the pinball excess risk of $\hat q_r=q_{\eta,\Omega}(r)+\delta_r$ is $\tfrac12 f_{Y\mid S_{\Omega}=r}(q_{\eta,\Omega}(r))\,\delta_r^{2} +o(\delta_r^{2})$ as $\delta_r\to0$, so Proposition~\ref{prop:cap} characterizes the leading local bias--variance behavior of the estimation risk. The cap sweep of Sec.~\ref{ssec:discussion} evaluates its predicted dependence on $K$. A small $K$ purges stale observations quickly, but the capped buffer
leaves a variance floor $\kappa_{\eta}\sigma_{1}^{2}/(\rho_{u_r}K)$ in
\eqref{eq:mse_cap} that shows up as worst-run instability. A large $K$ lowers the floor but lets obsolete samples persist for up to $K$ visits. The squared-bias term in \eqref{eq:mse_cap} does not distinguish the direction of drift, but the allocation consequences do. After upward drift, stale low-regime samples tend to cost coverage, whereas after downward drift, stale high-regime samples mainly cost resources. This is why Sec.~\ref{ssec:discussion} favors a smaller $K$ on the upward ramp and a larger one on the downward departure. No single $K$ suits both, so the cap is selected by worst-run coverage on a pilot set disjoint from the evaluation realizations. The representation term is independent of $K$, so the resource gain under representation-changing drift comes from re-deriving $\Omega$, while re-binning preserves the tail-estimation support.

\begin{figure}[t]
\centering
\vspace{-8pt}
\includegraphics[width=\columnwidth]{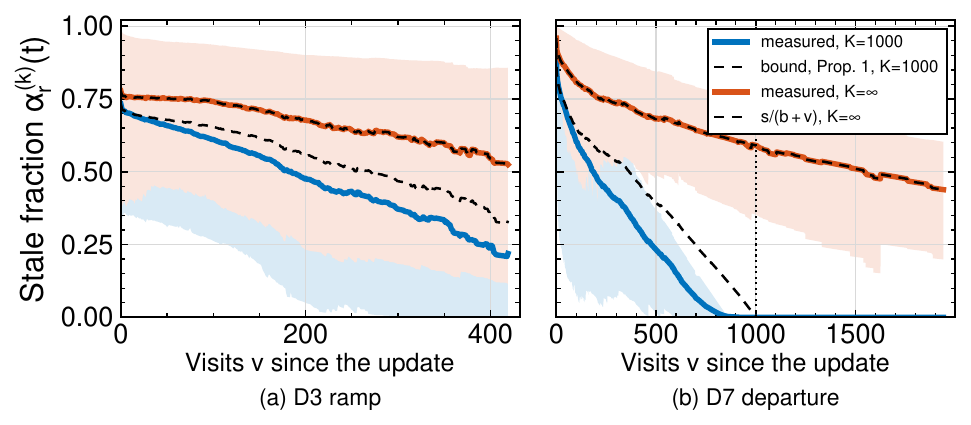}
\vspace{-14pt}
\caption{Measured stale fraction after an epoch update (mean over states and realizations, band: 10th--90th percentile) against the bound \eqref{eq:alpha_cap} at $K\!=\!10^{3}$ and the unbounded $s/(b+v)$ (dashed, indices dropped). Curves drawn where at least $20$ traces reach the visit count; dotted line, $v\!=\!K$.}
\label{fig:alpha}
\vspace{-6pt}
\end{figure}

Fig.~\ref{fig:alpha} tests the bound \eqref{eq:alpha_cap} against the
measured buffers for the two drift directions, an upward ramp and a
downward departure (D3 and D7 of Sec.~\ref{sec:results}), over $100$
realizations with traces ending at the next accepted epoch. Under the
cap, the measured stale fraction lies below the bound at every visit
and, on the departure, reaches zero at $v\!=\!923$, before the bound's
$v\!=\!K$. Without the cap, it follows the unbounded fraction of
Proposition~\ref{prop:cap} exactly and remains at $0.43$ after
$2{,}000$ visits.

\vspace{-6pt}
\subsection{LMRD-based threshold selection and tail refit}\label{ssec:lmrd}
The exceedance threshold marks where the GPD tail begins, a property of the current interference distribution, so a percentile $\hat q$ (Sec.~\ref{ssec:mixture}) chosen on the training regime need not remain valid after a change, and we re-select it at each epoch update. We choose a common percentile $\hat q$ from a finite
candidate set on Hosking's L-moment-ratio diagram (LMRD)~\cite{hosking1990lmoments}, set the state-specific
thresholds $u_{i}\!=\!Q_{\mathcal{D}_{i}}(\hat q)$ on the retained per-state samples of Sec.~\ref{ssec:memory}, and estimate the GPD parameters above them with the same L-moment apparatus. Three samples play distinct roles in what follows: the pooled selection sample on which $\hat q$ is chosen, the per-state buffers $\mathcal{D}_i$ on which the tail is fitted, and the $n_{\mathrm{refit}}$ new visits that trigger a per-state refit between epoch updates.

We compute L-moments via PWMs~\cite{landwehr1979pwm}. With $X_{1:n}\!\le\!\cdots\!\le\!X_{n:n}$ the order statistics of a sample, the unbiased PWM estimator of
$\beta_{r}\!=\!\mathbb{E}[X F(X)^{r}]$ is
\begin{equation}
\hat\beta_{r} \;=\; \frac{1}{n}\sum_{i=r+1}^{n}\!
\frac{\binom{i-1}{r}}{\binom{n-1}{r}}\,X_{i:n},
\quad r\!=\!0,\dots,3,
\label{eq:pwm}
\end{equation}
and the sample L-moments follow as $\hat\lambda_{1}\!=\!\hat\beta_{0}$, $\hat\lambda_{2}\!=\!2\hat\beta_{1}\!-\!\hat\beta_{0}$, $\hat\lambda_{3}\!=\!6\hat\beta_{2}\!-\!6\hat\beta_{1}\!+\!\hat\beta_{0}$,
$\hat\lambda_{4}\!=\!20\hat\beta_{3}\!-\!30\hat\beta_{2}\!+\!12\hat\beta_{1}\!-\!\hat\beta_{0}$, with the scale-free ratios $\hat\tau_{3}\!=\!\hat\lambda_{3}/\hat\lambda_{2}$ (L-skewness) and $\hat\tau_{4}\!=\!\hat\lambda_{4}/\hat\lambda_{2}$ (L-kurtosis). For the GPD \eqref{eq:gpd_cdf}, the population ratios are
\begin{equation}
\tau_{3} \!=\! \frac{1+\xi}{3-\xi},\qquad
\tau_{4} \!=\! \frac{(1+\xi)(2+\xi)}{(3-\xi)(4-\xi)},
\label{eq:gpd_locus}
\end{equation}
so as $\xi$ varies, $(\tau_{3},\tau_{4})$ traces a curve $\mathcal{C}_{\mathrm{GPD}}$, the \emph{GPD locus}, on which every GPD lies (Fig.~\ref{fig:pipeline}d). 
L-moment ratios are invariant under positive affine transformations, so the rules below are unchanged by linear gain and offset calibration of the measurement chain.

At each epoch update, on the post-anchor selection sample $\mathcal{S}^{(k)}(t)=\{I_r:\,a^{(k)}\!+\!1\le r\le t\}$, we select the threshold percentile as 
\begin{equation}
\hat q \;=\; \mathop{\arg\min}_{q\in\mathcal{U}}\;
\big\|\big(\hat\tau_{3}(q),\hat\tau_{4}(q)\big)
- \mathcal{C}_{\mathrm{GPD}}\big\|,
\label{eq:lmrd_u}
\end{equation}
where $\mathcal{U}$ is a finite grid of candidate percentiles and $\hat\tau_{3}(q),\hat\tau_{4}(q)$ are computed from the samples of $\mathcal{S}^{(k)}(t)$ above its $q$-quantile, pooled over states; we write $\Delta_{\mathrm{sel}}(q)$ for the minimized distance. We pool only for this selection, since a single state holds too few exceedances for it (Sec.~\ref{ssec:mixture}). Selecting $\hat q$ on the post-anchor sample, common to all three memory policies, keeps the choice independent of the policy. Selecting it on the retained buffers instead changes coverage by at most $2\!\times\!10^{-4}$ (Sec.~\ref{ssec:lmrd_results}). We admit a candidate only if it leaves at least eight pooled exceedances above its $q$-quantile, $n(1\!-\!q)\!\ge\!8$ with $n\!=\!|\mathcal{S}^{(k)}(t)|$. A window shorter than $n_{\Omega}$ thus loses the highest percentiles first, and if none qualify, the inherited $q\!=\!0.97$ is kept. Between epoch updates, $\hat q$ is held and only the per-state fits are refreshed.
\begin{algorithm}[t]
\small
\caption{Adaptive interference prediction}
\label{alg:main}
\begin{algorithmic}[1]
\STATE \textbf{Train} $\Omega^{(0)}$ and $\tilde\Theta_0$ on the stationary
       window $\mathcal{W}_{0}$; build buffers $\{\mathcal{D}_i\}$
       from $\mathcal{W}_0$; call
       \textsc{Refit}($\mathcal{W}_{0},\{\mathcal{D}_i\}$)
       to fit the initial mixture
\STATE \ \ $k\!\leftarrow\!0$;
       $a^{(0)}\!\leftarrow\!0$;\ $v_i\!\leftarrow\!0\ \forall i$
\FOR{each slot $t$}
  \STATE $s_t\!\leftarrow\!\ell(I_t;\Omega^{(k)})$ \eqref{eq:statemap};
         predict $\Ihat$ \eqref{eq:evt}; allocate $M$ \eqref{eq:M}
  \STATE append $I_t$ to $\mathcal{D}_{s_{t-1}}$, dropping its oldest
         member if $|\mathcal{D}_{s_{t-1}}|\!>\!K$;\
         $v_{s_{t-1}}\!\leftarrow\!v_{s_{t-1}}\!+\!1$
  \STATE update CUSUM \eqref{eq:cusum}; on crossing, confirm
         \eqref{eq:chi2}: accept past refractory $\Rightarrow$
         $\hat t^\ast$ \eqref{eq:cp}; reject or within refractory
         $\Rightarrow$ $m_i\!\leftarrow\!0$
  \IF{change accepted past refractory \eqref{eq:epoch}}
    \STATE $k\!\leftarrow\!k\!+\!1$;\ $a^{(k)}\!\leftarrow\!\hat t^\ast$
    \STATE re-derive $\Omega^{(k)}$ \eqref{eq:adaptomega};\ reset
           $\tilde\Theta_0$ on $\Omega^{(k)}$-discretized
           $[a^{(k)}\!+\!1,t]$
    \STATE re-bin the retained buffers $\{\mathcal{D}_i\}$ under
           $\Omega^{(k)}$ by each sample's stored predecessor; rebuild
           each $\mathcal{D}_i$ from its $K$ most recent members;\
           $v_i\!\leftarrow\!0\ \forall i$
    \STATE call \textsc{Refit}($\mathcal{S}^{(k)}(t),\{\mathcal{D}_i\}$)
           \quad{\small\textit{// Alg.~\ref{alg:refit}}}
  \ELSIF{$v_{s_{t-1}}\!\ge\!n_{\mathrm{refit}}$}
    \STATE refit state $s_{t-1}$ from $\mathcal{D}_{s_{t-1}}$ at the
           held $\hat q$ (lines 4--10 of Alg.~\ref{alg:refit}), which
           re-evaluates $\textit{fallback}_{s_{t-1}}$;\
           $v_{s_{t-1}}\!\leftarrow\!0$
  \ENDIF
\ENDFOR
\end{algorithmic}
\end{algorithm}

Given $\hat q$, we set the per-state thresholds $\hat u_{i}\!=\!Q_{\mathcal{D}_{i}}(\hat q)$ and estimate the GPD shape and scale with Hosking's closed-form L-moment estimator~\cite{hosking1990lmoments},
\begin{equation}
\hat\xi \;=\; \frac{3\hat\tau_{3}-1}{1+\hat\tau_{3}}, \qquad
\hat\sigma \;=\; \hat\lambda_{2}\,(1-\hat\xi)(2-\hat\xi),
\label{eq:lmom_gpd}
\end{equation}
computed from the L-moments of each state's exceedances. This replaces the iterative maximum-likelihood fit of~\cite{salehi2025evt} with the same L-moment machinery as \eqref{eq:lmrd_u}, and we retain the clip
$\hat\xi\in[-0.5,0.45]$ for finite-quantile stability. The estimator suits the small counts a state holds just after a re-bin (Sec.~\ref{ssec:lmrd_results}). It is closed-form, so a refit always returns. Also, the L-moments are linear in the order statistics, so an extreme observation enters with weight $O(1/n)$ rather than the leverage it gains in higher-order moments. Per state, the GPD is fitted only when the sample and exceedance gates of Algorithm~\ref{alg:refit} (lines 4--8) are met; otherwise the state is served by the smoothed empirical quantile, or by the upper bin edge when empty. 
With bounded per-state buffers of up to $K$ samples, $90$--$95\%$ of fits pass, the GPD branch serves $99.3$--$99.7\%$ of occupancy on every scenario and the KDE fallback under $1\%$, so the tail model operates where the predictions are made.
Algorithm~\ref{alg:refit} lists the complete refit.

The distance of \eqref{eq:lmrd_u} also serves as a defensive diagnostic. Per state,
\begin{equation}
\Delta_{\mathrm{GPD},i} \;=\; \big\|(\hat\tau_{3},\hat\tau_{4})
- \mathcal{C}_{\mathrm{GPD}}\big\|
\label{eq:lmrd_validate}
\end{equation}
is computed on the exceedances of $\mathcal{D}_i$ at every fit, epoch or periodic. Above a tolerance $\tau_{\mathrm{LMRD}}$ calibrated on stationary traces, the fitted tail is too far from GPD for \eqref{eq:evt} to be trusted, and state $i$ is served the empirical conditional quantile $Q_{\mathcal{D}_{i}}(\eta)$ until a later refit returns within tolerance. 
Immediately after a re-bin, state buffers are smallest and their LMRD distances noisiest, and $12$--$18\%$ of these fits exceed $\tau_{\mathrm{LMRD}}$ and fall back to the empirical quantile in every scenario, with or without drift.
\begin{algorithm}[t]
\small
\caption{\textsc{Refit}($\mathcal{S},\{\mathcal{D}_i\}$): LMRD threshold selection
and L-moment tail fit}
\label{alg:refit}
\begin{algorithmic}[1]
\REQUIRE selection sample
$\mathcal{S}$ (pooled); state buffers $\{\mathcal{D}_i\}$;
grid $\mathcal{U}$ (Table~\ref{tab:params});
tolerance $\tau_{\mathrm{LMRD}}$
\STATE for each \emph{admissible} $q\!\in\!\mathcal{U}$ ($\ge\!8$
       pooled exceedances of $\mathcal{S}$):
       $(\hat\tau_{3}(q),\hat\tau_{4}(q))$ from PWMs \eqref{eq:pwm}
\STATE $\hat q \leftarrow
       \arg\min_{q}\big\|\big(\hat\tau_{3}(q),\hat\tau_{4}(q)\big)
       \!-\!\mathcal{C}_{\mathrm{GPD}}\big\|$ \eqref{eq:lmrd_u}
\FOR{each state $i$ with buffer $\mathcal{D}_{i}$}
  \IF{$|\mathcal{D}_{i}|\!\ge\!5$ and $\ge\!5$ exceedances above
      $\hat u_{i}\!=\!Q_{\mathcal{D}_{i}}(\hat q)$}
    \STATE fit $(\hat\xi_{i},\hat\sigma_{i})$ \eqref{eq:lmom_gpd},
           clip $\hat\xi_{i}\!\in\![-0.5,0.45]$; KDE bulk below
           $\hat u_{i}$; $\rho_{u_i}\!\leftarrow\!$ exceedance
           fraction
    \STATE $\Delta_{\mathrm{GPD},i}\!\leftarrow\!$ \eqref{eq:lmrd_validate}
           on the exceedances of $\mathcal{D}_{i}$;
           $\textit{fallback}_i\!\leftarrow\!
           [\Delta_{\mathrm{GPD},i}\!>\!\tau_{\mathrm{LMRD}}]$
  \ELSE
    \STATE serve state $i$ by the smoothed empirical
           $\eta$-quantile of $\mathcal{D}_{i}$, or the upper bin
           edge if $\mathcal{D}_{i}$ is empty
  \ENDIF
\ENDFOR
\STATE at prediction, a state with $\textit{fallback}_i$ set is served
       $Q_{\mathcal{D}_{i}}(\eta)$ instead of \eqref{eq:evt}
\end{algorithmic}
\end{algorithm}

\subsection{Computational complexity}\label{ssec:complexity}

Algorithm~\ref{alg:main} runs every slot, so its online cost must be negligible against a URLLC slot. Per slot, state assignment costs $O(\log L)$ and the buffer, count and CUSUM updates $O(1)$. Re-estimating $\tilde\Theta_1$, scoring the block and confirming recur every $W$ slots, $O(L^{3}/W)$ per slot on average. A
per-state refit sorts at most $K$ samples once per $n_{\mathrm{refit}}$ visits, $O(K\log K/n_{\mathrm{refit}})$ per slot on average, and an epoch update re-bins and refits every state, $O(KL\log K)$, only at accepted alarms. The average per-slot cost is therefore $O(L^{3}/W+K\log K/n_{\mathrm{refit}})$,
about $10^{3}$ operations at the operating point and
independent of $t$; execution time is implementation- and platform-dependent. The epoch update can run in the background, the previous fit remaining active until the update completes. Memory is bounded regardless of run length. The predictor holds $KL$ sample pairs, the detector its $L_{\mathrm{det}}$-slot window, and the epoch update a post-anchor FIFO of $L_{\mathrm{win}}$ samples (the anchored baseline's window length, Sec.~\ref{ssec:memory}) from which $\Omega^{(k)}$, $\hat q$ and $\tilde\Theta_0$ are derived. Cumulative refitting instead needs $O(t)$ storage and $O(t\log t)$ refits. Re-binning these buffers rather than the full history changes coverage by at most $0.0008$ and resource use by at most $0.1\%$ on every scenario.

\section{Numerical Results}\label{sec:results}

We evaluate the proposed method by Monte Carlo simulation of the interference model in \eqref{eq:I}. Table~\ref{tab:params} lists the system, predictor, detector, and adaptation parameters. The detector, memory, and refit parameters are fixed a priori, without tuning on the evaluation realizations (their sensitivity is examined in Sec.~\ref{ssec:discussion}), and $\tau_{\mathrm{LMRD}}$ is calibrated
on stationary traces with seeds disjoint from evaluation. We verify all results on $100$ ($20$ for sensitivity analysis) independent realizations of $20000$ samples per scenario, reported with $95\%$ normal-approximation CIs over the run-level estimates.

\emph{Scenarios.} We evaluate drift resilience against eight scenarios based on distinct behavior of \eqref{eq:I}, six with the common drift onset $t^{\ast}\!=\!10^{4}$. Training uses the Table~\ref{tab:params} baseline ($N\!=\!5$, $\mu\!=\!0.4$) except where stated: \textbf{D1} \emph{interferers join} ($N\!:\!5\!\rightarrow\!10$, abrupt at $t^{\ast}$); \textbf{D2} \emph{load shift} ($\mu\!:\!0.4\!\rightarrow\!1.0$, abrupt at $t^{\ast}$); \textbf{D3} \emph{incremental ramp} ($3\times$
rescaling of all per-interferer INRs, linear over $t^{\ast}$ to $1.5\!\times\!10^{4}$); \textbf{D4} \emph{recurrent load} ($\mu$ alternating $1.0/0.4$ every $\approx\!3.3\!\times\!10^{3}$ samples
from $t^{\ast}$); \textbf{D5} \emph{stationary control} (no drift); \textbf{D6} \emph{compound} (stacked run of load shift at $t_{1}\!=\!5\!\times\!10^{3}$, the join at $t_{2}\!=\!10^{4}$, and a $\times2$ INR ramp over $3.3\!\times\!10^{3}$ samples from $t_{3}\!=\!1.5\!\times\!10^{4}$); \textbf{D7} \emph{departure}
($N\!:\!10\!\rightarrow\!5$ at $t^{\ast}$, trained at $N\!=\!10$, the downward direction); and \textbf{D8} \emph{scale reversal} (all INRs $\times3$ at $5\!\times\!10^{3}$, back at $10^{4}$, $\times3$ again at $1.5\!\times\!10^{4}$, so that the training partition is correct again on the return leg). D1, D3, D6, D7 and D8 shift the level distribution enough to move the quantile edges, the \emph{representation-changing} scenarios; D2 and D4 alter activity and transition dynamics but leave the partition essentially unchanged. All methods see the same realizations, initialization, and drift times, so every comparison is paired.
\begin{table}[t]
\vspace{-8pt}
\caption{Simulation and method parameters~\cite{salehi2025evt}.}
\vspace{-8pt}
\label{tab:params}
\centering
\small
\resizebox{0.8\linewidth}{!}{
\begin{tabular}{@{}llll@{}}
\toprule
\multicolumn{2}{@{}l}{System and predictor} &
\multicolumn{2}{l@{}}{Detector and adaptation}\\
\midrule
SNR $\bar\gamma$ & $20$~dB & subsequence $W$ & $15$\\
mean INR & $\mathcal{U}[-10,5]$~dB & detector window $L_{\mathrm{det}}$ & $300$\\
payload $b$ & $50$~bits & threshold $h$, level $\alpha$ & $10$, $0.05$\\
interferers $N$ & $5$ & refractory $\rho$, Laplace $\beta$ & $600$, $1$\\
levels $L$ & $15$ & memory cap $K$ & $10^{3}$/state\\
percentile grid $\mathcal{U}$ & $\{.93,.95,.97,.98\}$ & refit cadence $n_{\mathrm{refit}}$ & $20$ visits\\
duration $\zeta$ & $10$ slots & re-derivation gate $n_{\Omega}$ & $400$ slots\\
training & $10^{3}$ samples & passive window $L_w$ & $300$ slots\\
$(\eta,\epsilon)$ & $(0.99,10^{-5})$ & anchored/FIFO window $L_{\mathrm{win}}$ & $2000$ slots\\
max. blocklength $M_{\max}$ & $10^{5}$ & refit period & $200$ slots\\
\bottomrule
\end{tabular}
}
\end{table}

\emph{Baselines.} We compare against five references. The
\textit{genie} uses the realized next-slot interference $I_{t+1}$ and provides the resource lower bound; \textit{static} is the trained EVT/KDE predictor of~\cite{salehi2025evt} with neither refit nor re-derivation; \textit{passive} is a periodic sliding-window refit without change detection; \textit{cumulative}
is the periodic refit prescribed in~\cite{salehi2025evt}, on every past observation with the training $\Omega$; and \textit{cdm-$\Omega$} re-derives $\Omega$ at each detected change but discards the pre-change history (the anchored window of Sec.~\ref{sec:method}-D). The proposed method follows Algorithm~\ref{alg:main} with $K\!=\!10^{3}$, while its unbounded variant ($K\!=\!\infty$) isolates the cap. For a fair comparison, all methods use the same tail machinery of Sec.~\ref{ssec:lmrd}. Coverage is asymmetric, so it is reported together with the resource ratio RU/genie and, at fixed $\eta$, with the pinball loss $\psi_{\eta}(I-\Ihat)$ (Sec.~\ref{sec:method}-B) averaged over slots, which penalizes over-prediction at rate $1-\eta$ and under-prediction at rate $\eta$.

\vspace{-6pt}
\subsection{Coverage and resource cost under drift}
\label{ssec:recovery}
\begin{figure}[!t]
    \centering
  \subfloat[ D1: abrupt interferer join]{%
       \includegraphics[width=0.9\linewidth]{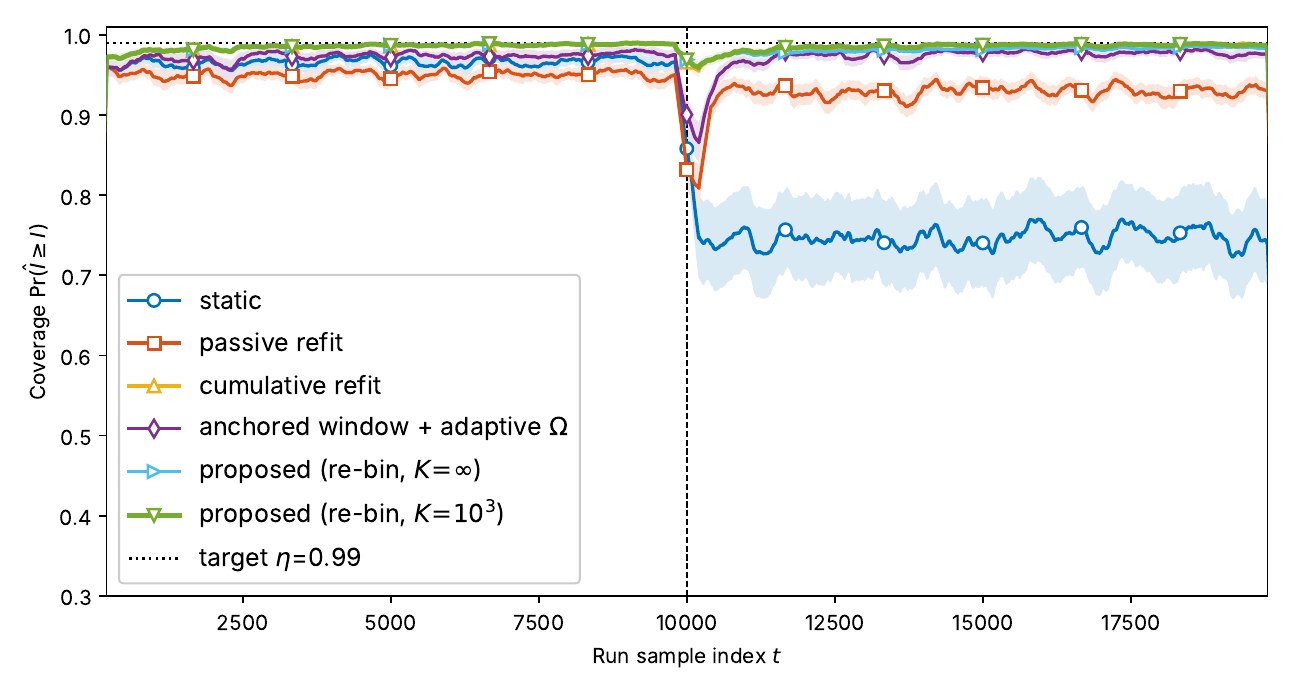}\label{fig:bands-a}}\\[-0.1em]
  \subfloat[D8: scale reversal, regime A
  (training) $\to$ B $\to$ A $\to$ B.]{%
        \includegraphics[width=0.9\linewidth]{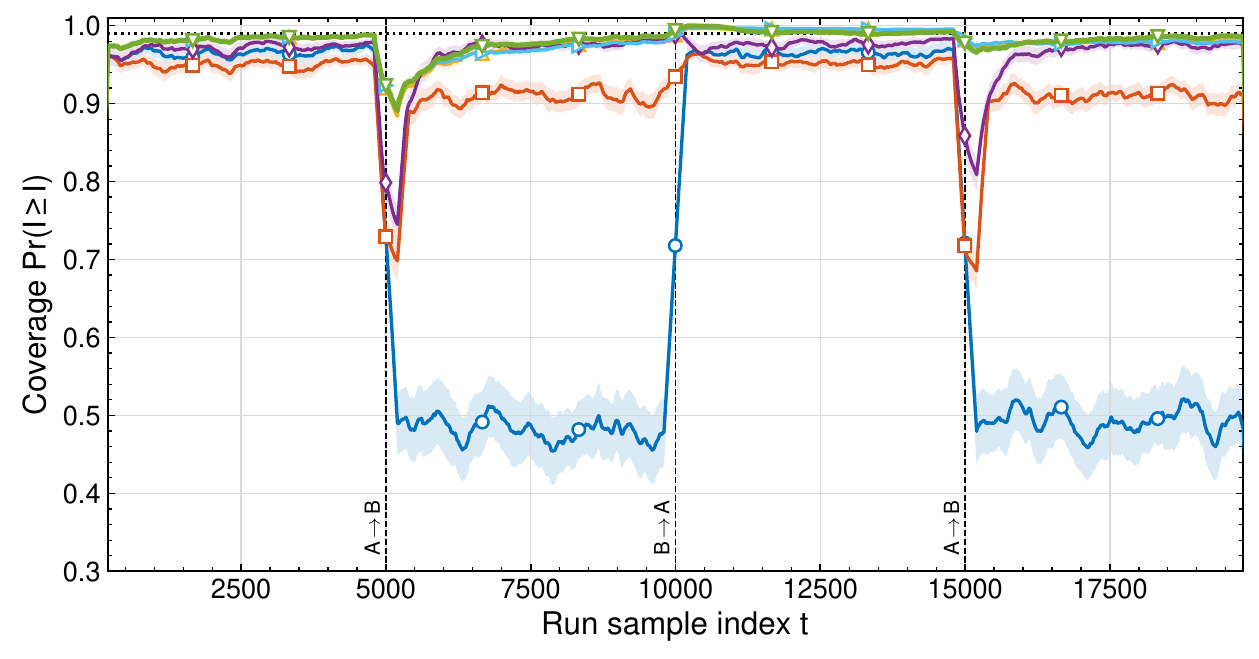}\label{fig:bands-b}}\\[-1pt]
  \caption{Coverage over time, $\pm$\,95\%\,CI.  The anchored window dips at every upward transition; the methods that retain history dip less than half as deep. Returning to A causes no material coverage loss.}
  \label{fig:bands}
  \vspace{-8pt}
\end{figure}
We first examine coverage over time (Fig.~\ref{fig:bands}) and the settled post-drift values (Table~\ref{tab:main}). Two comparisons isolate the two design choices: cumulative versus proposed shows the effect of re-deriving the state space, and cdm-$\Omega$ versus proposed shows the effect of retaining history.

Under abrupt D1 (Fig.~\ref{fig:bands}(a)), static coverage drops at the onset and never recovers, passive refitting recovers only partially, and the cumulative and adaptive methods return near their pre-drift levels.
The transient separates them: the anchored window dips to $0.87$, whereas the proposed method and cumulative refit, both retain history, dip only to $0.96$. The reversal D8 (Fig.~\ref{fig:bands}(b)) repeats this at each A$\to$B transition ($0.75$ against $0.89$ at the first), while the return B$\to$A causes no material coverage loss, since the training partition is correct again. Discarding history is thus a transient cost at every upward change, and re-binning retained history removes most of it.

Detection delay, shared by every detection-driven method, depends on how fast and how far a scenario shifts the level distribution. Median delays are $200$--$370$ samples for the abrupt D1, D7 and D8, against $1{,}500$--$1{,}850$ for the ramp D3 and the load-only D2 and D4. In D2 and D4, about a quarter of the runs miss the change altogether, and the medians are over the runs that detect it. D6 is bimodal: about half the runs detect the initial mild load shift, with a median delay near $1{,}000$ samples, and the rest first fire at the join $5{,}000$ samples later (Fig.~\ref{fig:S-recov}, Appendix~\ref{sec:S-delay}).

Table~\ref{tab:main} gives post-transient coverage, resource use, and pinball loss. Static collapses at near- or below-genie resource use, and passive recovers only partly. 
Against cumulative refit, the proposed method matches or raises coverage on every representation-changing
scenario but D7 while lowering resource use by $6$--$13\%$ and pinball loss by $25$--$50\%$; with common realizations and the same tail estimator, the difference is the state partition and memory policy. Against the
anchored window, it raises coverage by $1.1$--$1.5$ pp at similar resource use and lowers pinball by $38$--$47\%$, so retaining history buys accuracy at marginal channel-use cost. On the stationary D5 and the load-only D2 and D4, the three cumulative variants agree to the third decimal, although the detector accepts $\approx\!5$ false alarms per run on D5: a false alarm costs the anchored window its accumulated sample but costs the proposed method only a partition it replaces in kind. The departure D7 is the one concession, at $11\%$ fewer channel uses, since the old high regime retained by the unbounded store is conservative there and the cap purges it. On the compound D6, the proposed method keeps its gain at $9\%$ fewer channel uses.
\begin{table}[t]
\caption{Post-drift coverage ($95\%$ CI), violation rate, RU/genie and pinball loss at $\eta\!=\!0.99$.
Bold: best in the refitting methods.}
\vspace{-8pt}
\label{tab:main}
\centering
\small
\resizebox{0.8\linewidth}{!}{
\begin{tabular}{@{}lllll@{}}
\toprule
Scenario / method & coverage & viol. & RU/genie & pinball\\
\midrule
\textbf{D1} static & 0.746\,[0.699,0.794] & 0.254 & 1.05 & 0.895\\
\quad\, passive & 0.930\,[0.926,0.933] & 0.070 & 1.20 & 0.141\\
\quad\, cumulative & 0.985\,[0.984,0.986] & 0.015 & 1.27 & 0.048\\
\quad\, cdm-$\Omega$ (anchored) & 0.975\,[0.974,0.976] & 0.025 & \textbf{1.19} & 0.059\\
\quad\, \textbf{proposed} ($K\!=\!10^{3}$) & \textbf{0.986\,[0.985,0.987]} & 0.014 & 1.20 & \textbf{0.036}\\
\quad\, proposed ($K\!=\!\infty$) & 0.981\,[0.980,0.983] & 0.019 & \textbf{1.19} & 0.037\\
\midrule
\textbf{D3} static & 0.594\,[0.567,0.622] & 0.406 & 0.92 & 2.714\\
\quad\, passive & 0.912\,[0.909,0.914] & 0.088 & 1.32 & 0.364\\
\quad\, cumulative & 0.964\,[0.963,0.965] & 0.036 & 1.42 & 0.143\\
\quad\, cdm-$\Omega$ (anchored) & 0.962\,[0.960,0.964] & 0.038 & 1.23 & 0.148\\
\quad\, \textbf{proposed} ($K\!=\!10^{3}$) & \textbf{0.973\,[0.973,0.974]} & 0.027 & 1.24 & \textbf{0.080}\\
\quad\, proposed ($K\!=\!\infty$) & 0.964\,[0.963,0.965] & 0.036 & \textbf{1.22} & 0.084\\
\midrule
\textbf{D5} static & 0.965\,[0.960,0.971] & 0.035 & 1.18 & 0.069\\
\quad\, passive & 0.951\,[0.949,0.953] & 0.049 & 1.18 & 0.058\\
\quad\, cumulative & \textbf{0.987\,[0.987,0.987]} & 0.013 & 1.20 & \textbf{0.022}\\
\quad\, cdm-$\Omega$ (anchored) & 0.974\,[0.972,0.976] & 0.026 & \textbf{1.20} & 0.046\\
\quad\, \textbf{proposed} ($K\!=\!10^{3}$) & \textbf{0.987\,[0.987,0.988]} & 0.013 & 1.21 & \textbf{0.022}\\
\quad\, proposed ($K\!=\!\infty$) & \textbf{0.987\,[0.987,0.988]} & 0.013 & 1.21 & \textbf{0.022}\\
\midrule
\textbf{D6} static & 0.652\,[0.618,0.686] & 0.348 & 0.92 & 2.270\\
\quad\, passive & 0.923\,[0.920,0.926] & 0.077 & 1.28 & 0.222\\
\quad\, cumulative & 0.973\,[0.972,0.975] & 0.027 & 1.33 & 0.107\\
\quad\, cdm-$\Omega$ (anchored) & 0.966\,[0.964,0.967] & 0.034 & \textbf{1.20} & 0.097\\
\quad\, \textbf{proposed} ($K\!=\!10^{3}$) & \textbf{0.979\,[0.978,0.979]} & 0.021 & 1.20 & \textbf{0.054}\\
\quad\, proposed ($K\!=\!\infty$) & 0.974\,[0.973,0.975] & 0.026 & \textbf{1.20} & 0.056\\
\midrule
\textbf{D7} static & 0.991\,[0.989,0.993] & 0.009 & 1.43 & 0.039\\
\quad\, passive & 0.958\,[0.957,0.960] & 0.042 & 1.34 & 0.057\\
\quad\, cumulative & \textbf{0.994\,[0.993,0.994]} & 0.006 & 1.41 & 0.031\\
\quad\, cdm-$\Omega$ (anchored) & 0.976\,[0.974,0.977] & 0.024 & \textbf{1.24} & 0.040\\
\quad\, \textbf{proposed} ($K\!=\!10^{3}$) & 0.989\,[0.989,0.990] & 0.011 & 1.26 & \textbf{0.023}\\
\quad\, proposed ($K\!=\!\infty$) & 0.992\,[0.991,0.992] & 0.008 & 1.28 & \textbf{0.023}\\
\midrule
\textbf{D8} static & 0.651\,[0.630,0.672] & 0.349 & 0.93 & 2.505\\
\quad\, passive & 0.918\,[0.916,0.921] & 0.082 & 1.32 & 0.346\\
\quad\, cumulative & 0.981\,[0.981,0.982] & 0.019 & 1.52 & 0.121\\
\quad\, cdm-$\Omega$ (anchored) & 0.968\,[0.967,0.969] & 0.032 & \textbf{1.31} & 0.145\\
\quad\, \textbf{proposed} ($K\!=\!10^{3}$) & \textbf{0.983\,[0.983,0.984]} & 0.017 & 1.33 & \textbf{0.076}\\
\quad\, proposed ($K\!=\!\infty$) & 0.981\,[0.981,0.981] & 0.019 & 1.33 & \textbf{0.076}\\
\bottomrule
\end{tabular}
}
\vspace{-8pt}
\end{table}
\begin{table}[t]
\caption{D8 by regime, coverage/RU per genie, first $500$
samples after each boundary excluded.}
\vspace{-8pt}
\label{tab:d8}
\centering
\small
\resizebox{0.8\linewidth}{!}{
\begin{tabular}{@{}lcccc@{}}
\toprule
method & A$_1$ (train) & B$_1$ ($\times3$) & A$_2$ (return) & B$_2$ ($\times3$)\\
\midrule
static & 0.965/1.18 & 0.481/0.85 & 0.965/1.18 & 0.492/0.87\\
passive & 0.950/1.18 & 0.910/1.36 & 0.952/1.18 & 0.910/1.37\\
cumulative & 0.983/1.20 & 0.968/1.56 & 0.995/1.28 & 0.979/1.61\\
cdm-$\Omega$ (anchored) & 0.972/1.19 & 0.972/1.30 & 0.977/1.33 & 0.971/1.31\\
\textbf{proposed} ($K\!=\!10^{3}$) & 0.983/1.19 & 0.975/1.30 & 0.992/1.39 & 0.982/1.31\\
proposed ($K\!=\!\infty$) & 0.983/1.19 & 0.968/1.29 & 0.996/1.42 & 0.978/1.32\\
\bottomrule
\end{tabular}
}
\vspace{-6pt}
\end{table}
\begin{table}[!t]
\caption{Proposed vs cumulative refit at matched achieved
coverage (columns~2--4) and at $\eta\!=\!0.999$ (columns~5--7).
Bootstrap intervals: $2000$ paired resamples of the $100$
realizations.}
\vspace{-8pt}
\label{tab:frontier}
\centering
\small
\resizebox{0.95\linewidth}{!}{
\begin{tabular}{@{}lccccccc@{}}
\toprule
 & \multicolumn{3}{c}{matched coverage} & \multicolumn{3}{c}{$\eta\!=\!0.999$}\\
\cmidrule(lr){2-4}\cmidrule(lr){5-7}
 & common $\bar c$ & min.\ saving [95\% CI] & max.\ & $\Delta\bar c$ & RU saving & $\Delta$pinball\\
Scn. & interval & (\%) & (\%) & (pp) & (\%) & (\%)\\
\midrule
D1 & 0.944--0.996 & +3.7 [+2.2, +5.3] & +7.6 & -0.02 & +7.8 & +0.4\\
D2 & 0.946--0.997 & +0.0 [$-$0.4, +0.3] & +0.4 & +0.01 & -0.1 & +0.3\\
D3 & 0.922--0.988 & +10.8 [+9.5, +12.0] & +17.5 & +0.09 & +17.1 & -1.2\\
D4 & 0.948--0.997 & +0.0 [$-$0.3, +0.3] & +0.4 & +0.02 & -0.0 & -3.0\\
D5 & 0.949--0.996 & $-$0.4 [$-$0.6, $-$0.2] & $-$0.3 & +0.01 & -0.4 & -1.7\\
D6 & 0.932--0.991 & +6.4 [+5.3, +7.6] & +12.9 & -0.04 & +13.1 & -10.6\\
D7 & 0.971--0.998 & +8.4 [+6.7, +10.5] & +9.5 & -0.10 & +10.5 & -18.4\\
D8 & 0.941--0.995 & +8.3 [+7.0, +9.8] & +16.4 & -0.03 & +16.7 & -7.5\\
\bottomrule
\end{tabular}
}
\vspace{-6pt}
\end{table}

Table~\ref{tab:d8} breaks down the reversal D8 by regime. On the two B legs, the proposed method uses $17$--$19\%$ fewer resources than cumulative refit at higher coverage. On the return leg A$_2$ the order inverts:
cumulative is cheapest because its training $\Omega$ is correct again, whereas the proposed method re-derives a coarser partition from the post-anchor window and over-provisions until the epoch matures. Re-binning wins when the regime leaves the training partition and costs when it returns. Recognizing the regime and reloading its stored partition instead is a natural extension, left to future work.
\begin{figure*}[!t]
    \centering
  \subfloat[Achieved versus target outage post-drift on D3]{%
       \includegraphics[width=0.4\linewidth]{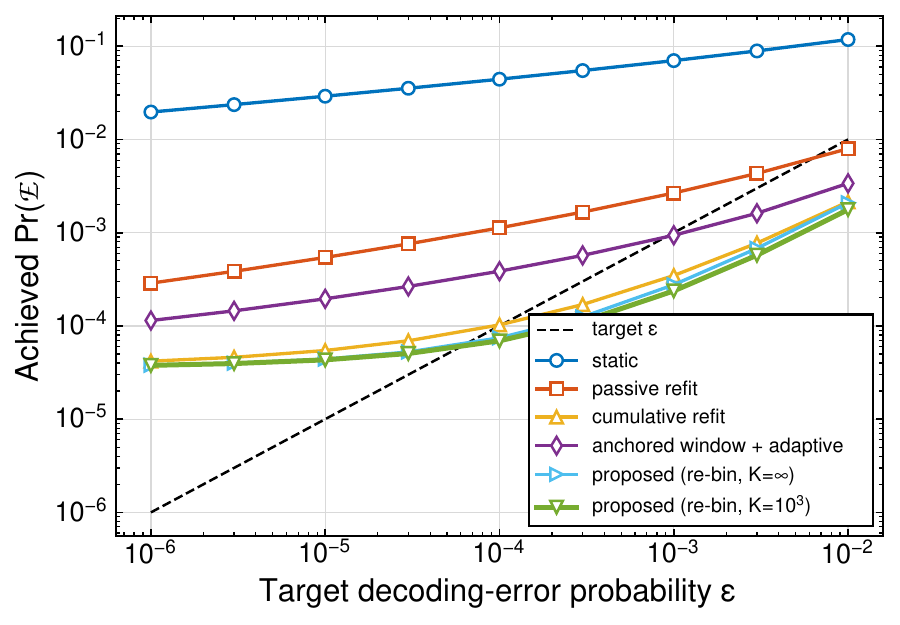}\label{fig:conv-a}}
  \subfloat[Reliability--resource plane at $\eta\!=\!0.99$]{%
        \includegraphics[width=0.4\linewidth]{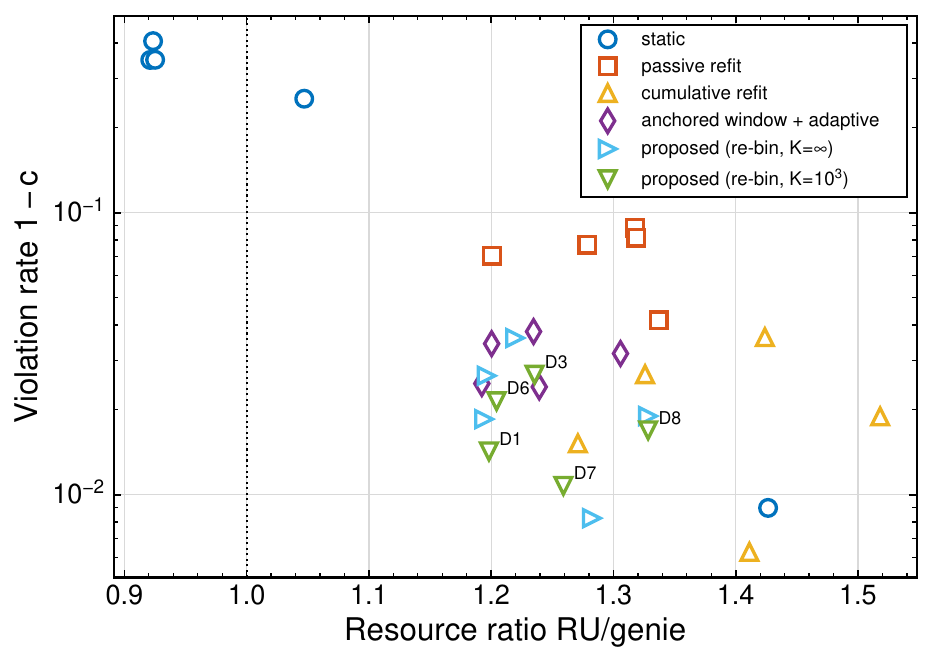}\label{fig:conv-b}}\\[-1pt]
  \caption{Post-drift reliability, $100$ realizations. In (b), one point per method--scenario pair, lower-left is better, and the dashed line marks RU/genie $=1$.}
\label{fig:conv}
\vspace{-18pt}
\end{figure*}

Since $\eta$ is a design knob that methods map to achieved coverage differently, Table~\ref{tab:frontier} compares at matched achieved coverage, interpolating each method's $(\bar c,\mathrm{RU})$ locus over the five $\eta$ levels. A paired bootstrap confirms the resource saving over the whole common interval on every representation-changing scenario: at least $2.2\%$ on D1 and $6$--$11\%$ on D3, D6, D7 and D8 at its
minimum, $8$--$18\%$ at its upper end. At matched coverage, D7's fixed-$\eta$ coverage deficit disappears, and its saving holds at $8.4$--$9.5\%$. On the load-only D2 and D4, the interval includes zero, and on the stationary D5 the method costs $0.4\%$, the measured price of running the detector where nothing changes. At $\eta\!=\!0.999$ coverage matches within $0.10$~pp, the resource saving widens to $8$--$17\%$ on the
representation-changing scenarios, and pinball stays lower on D3--D8. At equal achieved coverage, then, the proposed method saves resources wherever drift changes the state representation and costs at most $0.4\%$ where it does not, a safe default when drift is unknown in advance.

Fig.~\ref{fig:conv} shows post-drift reliability against the decoding-error target (a) and against resource use (b). In (a), static saturates nearly flat in $\epsilon$, so tightening the target buys no reliability, whereas the refitting methods track the target down to the floor of $\approx\!4\!\times\!10^{-5}$ set
by $1-\eta$; the proposed method has the lowest achieved outage at every $\epsilon$, $4.3\!\times\!10^{-5}$ against cumulative's $5.5\!\times\!10^{-5}$ and the anchored window's $1.9\!\times\!10^{-4}$ at the operating target, and the same holds on D1. In (b), at $\eta\!=\!0.99$ the proposed method lies left of cumulative refit on every representation-changing scenario and below it on all but D7, where it trades a small coverage
reduction for lower resource use (Table~\ref{tab:frontier}: $\approx\!9\%$ saving at matched coverage); it lies below the anchored window at the same resource use, while static's D3, D6 and D8 points fall in the under-provisioning quadrant.

\begin{figure}[t]
\centering
\subfloat[achieved coverage]{\includegraphics[width=0.5\columnwidth]{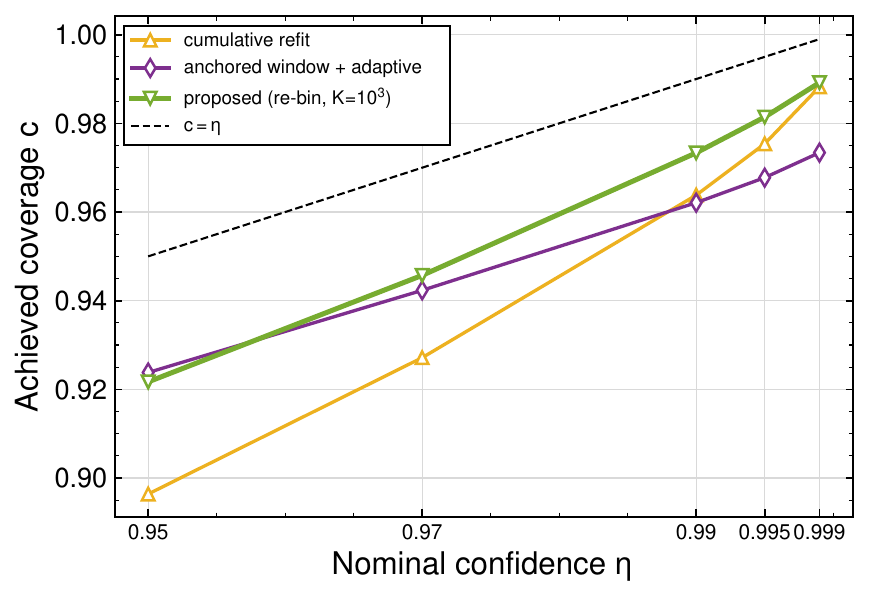}}\hfil
\subfloat[resource ratio]{\includegraphics[width=0.5\columnwidth]{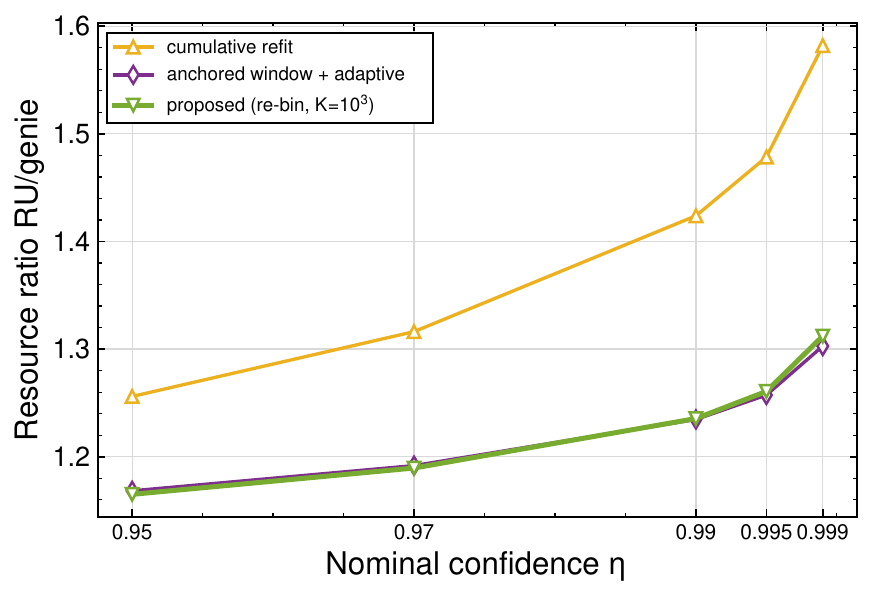}}
\vspace{-2pt}
\caption{Response to the nominal confidence $\eta$ on D3: (a)~coverage, where the anchored window trails and the proposed method follows $\eta$ above the cumulative refit; (b)~resource use, lower for the proposed method at every $\eta$.}
\label{fig:eta}
\vspace{-10pt}
\end{figure}

The same holds across the nominal confidence $\eta$. Over $\eta\in\{0.95,\dots,0.999\}$ the anchored window's coverage on D3,  where contamination lasts longest, trails the proposed method's by $2$--$3$ percentage points at every level (Fig.~\ref{fig:eta}(a)), and the proposed method's resource cost stays below the cumulative refit's (Fig.~\ref{fig:eta}(b)). This
includes $\eta\!=\!0.95$, below the exceedance threshold, where no tail model is in play, and the gain comes solely from re-binning.
\begin{figure}[!t]
\centering
\includegraphics[width=0.85\columnwidth]{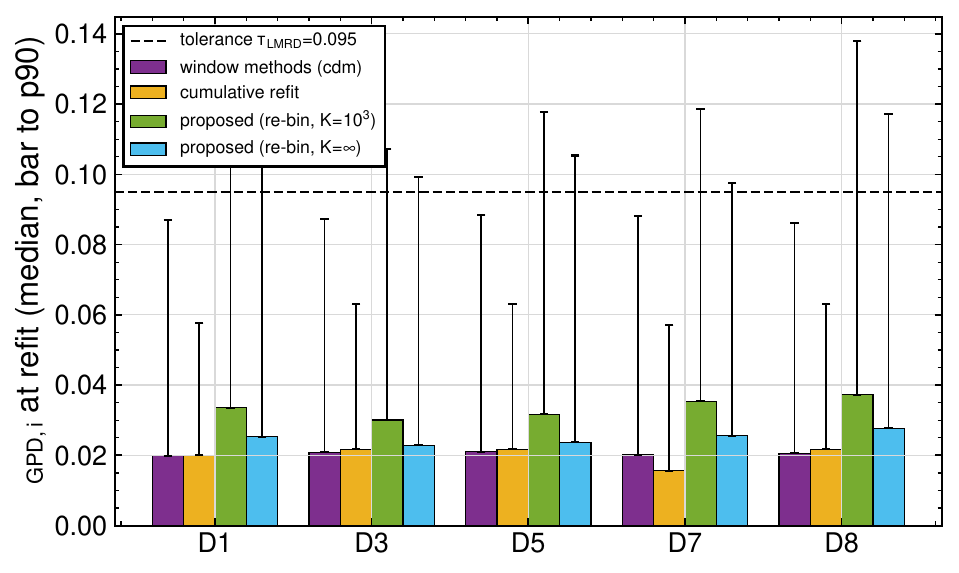}
\vspace{-10pt}
\caption{GPD-fitness diagnostic \eqref{eq:lmrd_validate} at epoch refits: median (bar) and $90$th percentile (whisker). Evaluated on the sample each refit fitted: the post-anchor window for the window-based methods,
the reconstructed buffers for unbounded retention, the per-state buffer for the proposed method.}
\label{fig:lmrdval}
\vspace{-6pt}
\end{figure}

\vspace{-10pt}
\subsection{Tail-refit validation}\label{ssec:lmrd_results}
We examine LMRD threshold-selection rule \eqref{eq:lmrd_u} and per-state tail diagnostic \eqref{eq:lmrd_validate} separately. The rule selects the exceedance percentile $\hat q$ by minimizing the pooled L-moment distance $\Delta_{\mathrm{sel}}(q)$ to the GPD locus. On stationary D5 traces, this distance averages $0.028$ at $q\!=\!0.93$ against $0.041$ at the inherited $q\!=\!0.97$, a difference of $0.013\!\pm\!0.009$. However, a lower distance does not translate into a coverage gain: a fixed $q\!=\!0.97$ instead of the rule changes coverage by at most $3\!\times\!10^{-4}$, and selecting $\hat q$ on the retained buffers instead of the post-anchor sample changes it by at most $2\!\times\!10^{-4}$.
\begin{figure*}[t]
\centering
\subfloat[mean coverage]{\includegraphics[width=0.3\linewidth]{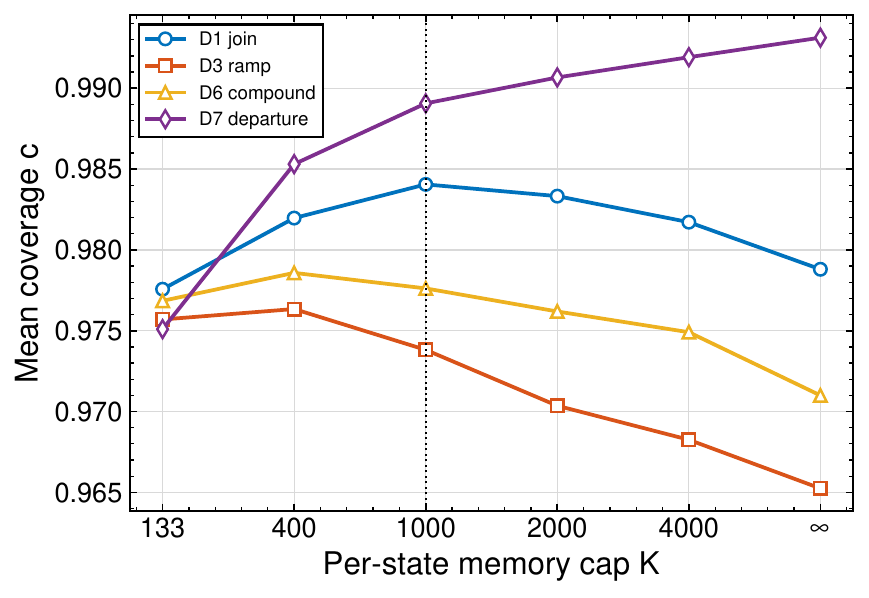}}\hfil
\subfloat[worst-run coverage]{\includegraphics[width=0.3\linewidth]{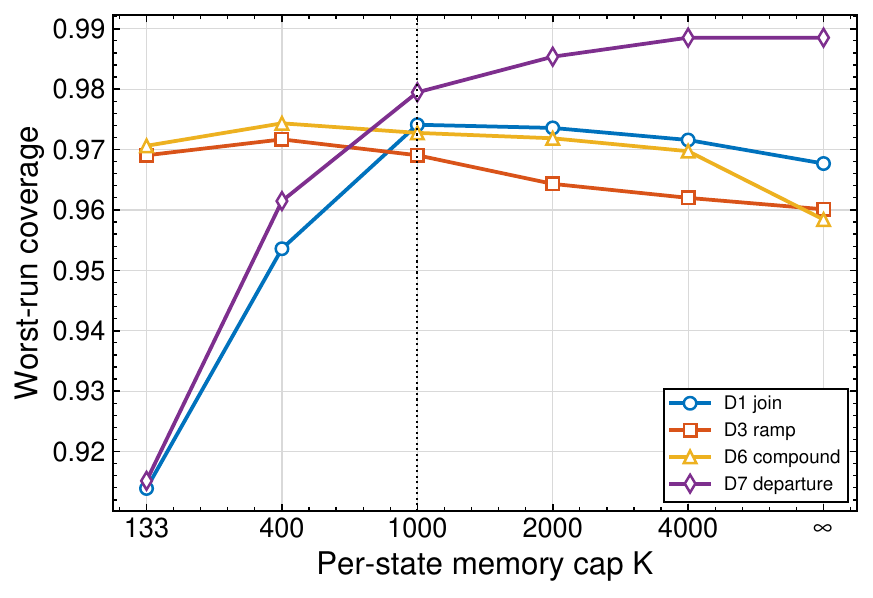}}\hfil
\subfloat[samples held]{\includegraphics[width=0.33\linewidth]{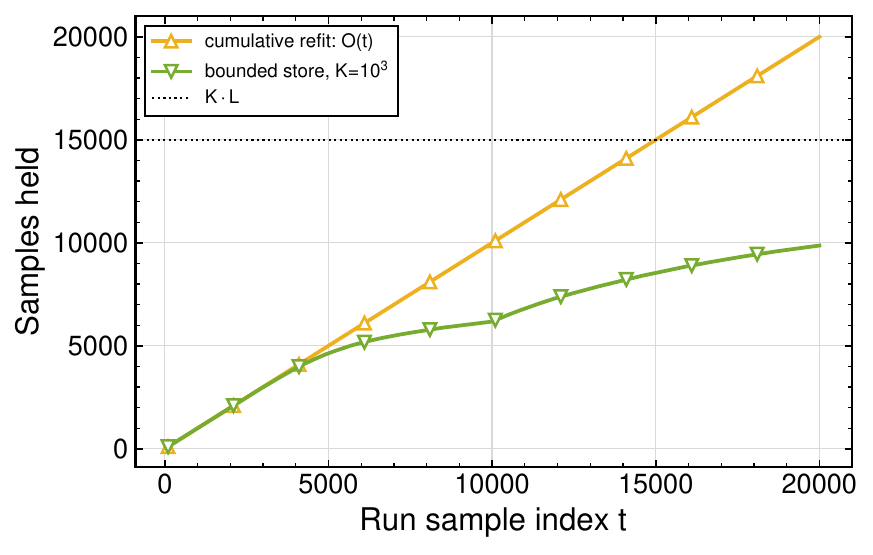}}
\vspace{-3pt}
\caption{Memory cap $K$ ($20$ realizations). The dotted line in (a), (b) marks $K\!=\!1000$, the smallest cap with a worst run above $0.96$ everywhere. In (c), unbounded retention grows as $O(t)$ while the bounded store saturates below $K\!\cdot\!L$.}
\label{fig:cap}
\vspace{-14pt}
\end{figure*}

The per-state diagnostic $\Delta_{\mathrm{GPD},i}$ instead checks whether a fitted tail is consistent with the GPD model. A fit exceeding the tolerance $\tau_{\mathrm{LMRD}}\!=\!0.095$ falls back to the empirical quantile. We set the tolerance to the $95$th percentile of the pooled selection distance on independent $2000$-sample windows of the stationary control D5 at $q\!=\!0.97$. This is a stationary screening reference, not a calibrated $5\%$ per-state rejection rate. Each window holds about $60$ exceedances, comparable to the $20$--$70$ in a full $K\!=\!10^{3}$ state buffer; smaller post-re-bin buffers give noisier diagnostics. Fig.~\ref{fig:lmrdval} shows median distances of $0.016$--$0.028$ for the window-based, cumulative and unbounded-retention refits and $0.030$--$0.037$ for the proposed bounded buffers. All medians stay below the tolerance, but the $90$th percentile exceeds it for both re-binning variants in every scenario, and the empirical fallback serves $12$--$18\%$ of the proposed method's epoch refits.

The estimator and the shape clip contribute separately. On $n\!=\!15$ i.i.d.\ GPD exceedances, the L-moment fit lowers the $\hat\xi$ MSE by $35\%$ over the maximum-likelihood fit of~\cite{salehi2025evt} and halves the $\hat\sigma$ MSE, a small-sample accuracy gain. Stability at tight targets comes from the clip: at $\eta\!=\!0.999$ the unclipped fit of~\cite{salehi2025evt} reaches $\hat\xi\!=\!3.7$ and predictions of $4\!\times\!10^{7}$ on the stationary control, whereas the same fit clipped to $\hat\xi\!\in\![-0.5,0.45]$ and the clipped L-moment fit agree in coverage within $0.001$.

The statewise gates decide where the GPD branch operates. For window-based refits, only $\approx\!20\%$ of state fits pass the five-exceedance gate, carrying $75$--$86\%$ of occupancy. With bounded per-state buffers, on every scenario $90$--$95\%$ of fits pass, the GPD branch serves $99.3$--$99.7\%$ of occupancy and the KDE-only states under $1\%$, so the tail model operates where the predictions are made.

\vspace{-12pt}
\subsection{Sensitivity, Robustness, and Scope}\label{ssec:discussion}
\begin{figure}[!t]
\centering
\vspace{-8pt}
\subfloat[D1 post-drift coverage]{\includegraphics[width=0.4\columnwidth]{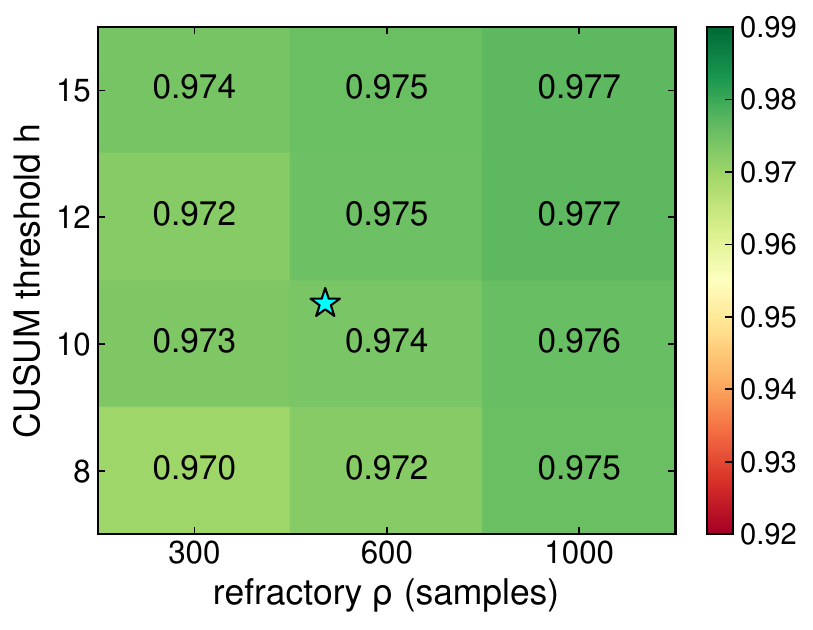}}\hfil
\subfloat[D1 RU/genie]{\includegraphics[width=0.4\columnwidth]{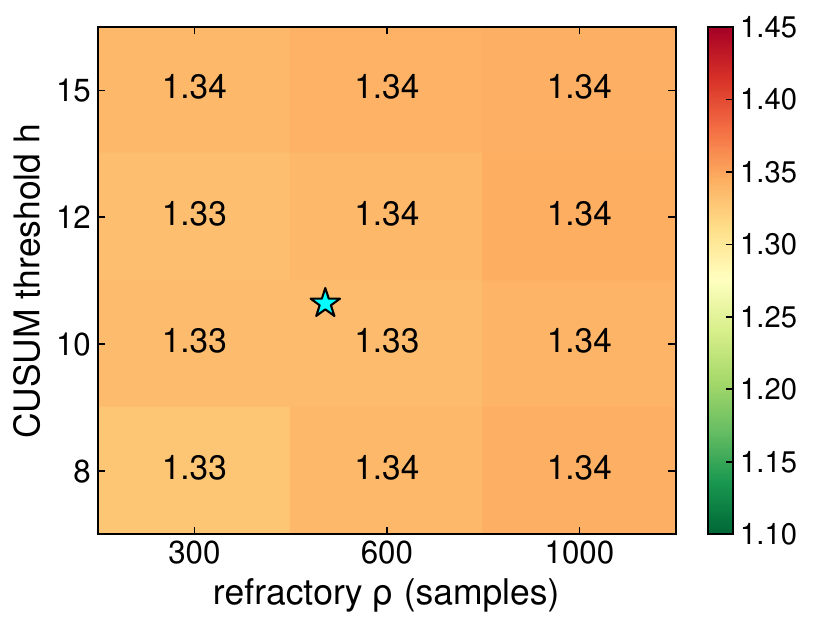}}\\[-10pt]
\subfloat[D1 detection delay]{\includegraphics[width=0.4\columnwidth]{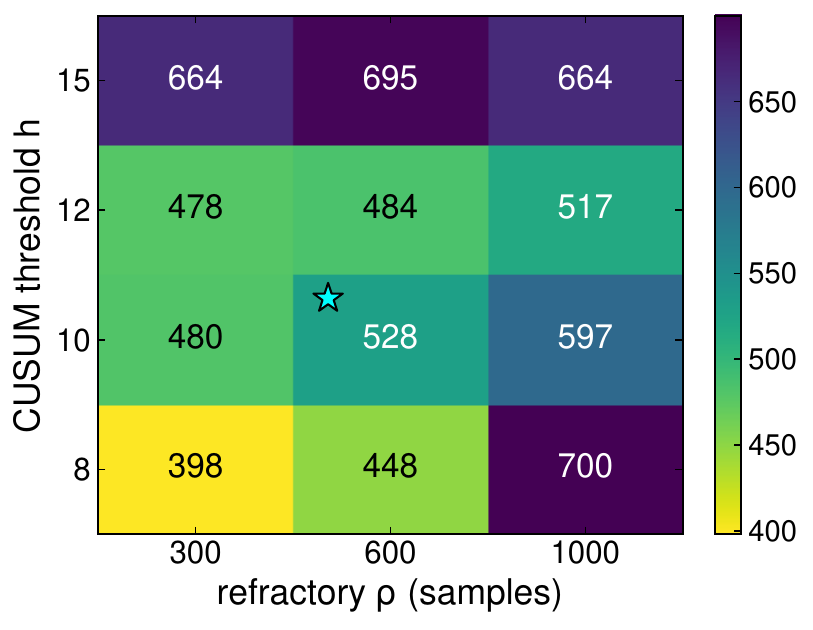}}\hfil
\subfloat[D5 false alarms per run]{\includegraphics[width=0.4\columnwidth]{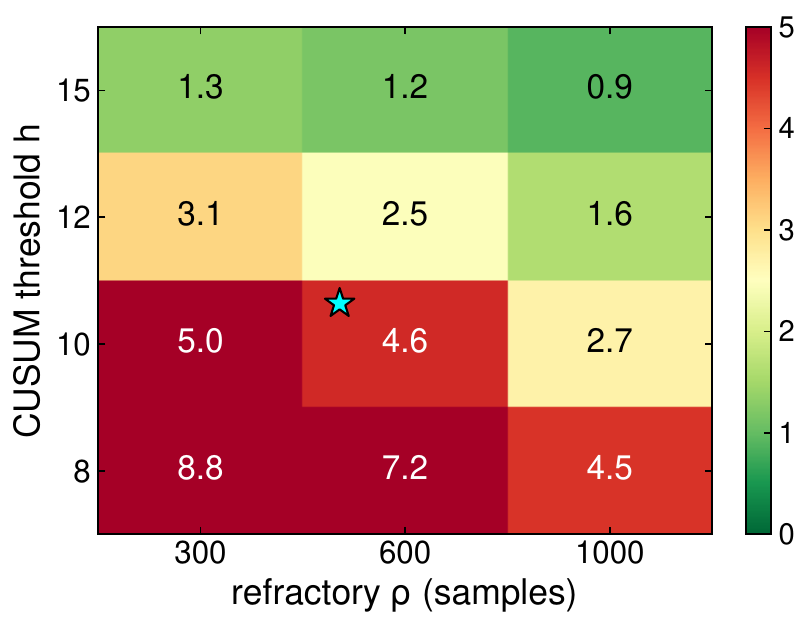}}
\vspace{-2pt}
\caption{Sensitivity of the detector to $(h,\rho)$, measured on the change-point-anchored window with the training $\Omega$. Star: operating point $(10,600)$.}
\label{fig:ablation}
\end{figure}

We now examine the sensitivity of results to the detector and the memory and refit parameters. The CUSUM threshold $h$ \eqref{eq:cusum} and refractory $\rho$ \eqref{eq:epoch} are swept jointly on the anchored window with the training $\Omega$ (Fig.~\ref{fig:ablation}), using D1, where a single abrupt onset makes detection delay well defined, and the stationary D5, where every accepted detection is a false alarm. Coverage and resource use are nearly invariant over the grid, spanning $[0.970,0.977]$ and $1.33$--$1.34$ (Fig.~\ref{fig:ablation}(a),(b)). Mean detection delay ranges over $400$--$700$ samples and tends to increase with $h$ and $\rho$, though not monotonically (Fig.~\ref{fig:ablation}(c)), while false alarms fall monotonically in both, from $8.8$ at $(8,300)$ to $0.9$ at $(15,1000)$ (Fig.~\ref{fig:ablation}(d)). The operating point therefore sits in a region of stable coverage and resource use rather than at a tuned optimum. The detector's trailing window is equally uncritical for coverage: over $L_{\mathrm{det}}\!\in\![150,1000]$ the proposed method's coverage moves by $0.001$ (Fig.~\ref{fig:windows}(b), Appendix~\ref{sec:S-windows}).

The last-zero anchor \eqref{eq:cp} is a retraining boundary, not a change-point estimate, but its accuracy sets what the post-anchor window contains. On the abrupt D1 and D7, the median $\hat t^{\ast}-t^{\ast}$ is $-10$ and $+73$ samples, and the median absolute error is $113$ and $128$, inside $L_{\mathrm{det}}\!=\!300$. The $5$th--$95$th percentiles reach $-550$ and $+1070$ on D7; the late tail is one run in ten, where the change lands in the refractory period of a preceding alarm and the first accepted epoch belongs to a later detection. At that detection, the post-anchor window reaches $n_{\Omega}$ in about one run in four, so most epochs open under $\Omega^{(k-1)}$ and re-derive at a later accepted event with enough samples.

The memory cap $K$ trades the two drift directions against each other (Fig.~\ref{fig:cap}(a)): coverage on the ramp D3 falls from $0.976$ at $K\!=\!133$ to $0.965$ unbounded, while on the departure D7 it rises from $0.975$ to $0.993$. 
Proposition~\ref{prop:cap} characterizes the bias--variance trade-off in $K$, and the drift direction determines whether stale samples primarily cost coverage or resources (Sec.~\ref{ssec:memory}). The worst run decides the choice (Fig.~\ref{fig:cap}(b)): $K\!=\!133$ falls to $0.91$ on D1 and D7, and $K\!=\!1000$ is the smallest cap with a worst run above $0.96$ on every scenario of a disjoint pilot set. The cap binds on about half the states, holding $9{,}900$ samples at $t\!=\!2\!\times\!10^{4}$ against $20{,}000$ unbounded (Fig.~\ref{fig:cap}(c)). The gate $n_{\Omega}$ is flat over $100$--$800$ ($0.972$--$0.975$ on D3); the cadence $n_{\mathrm{refit}}$ is not: at $200$ visits on D3 both retention variants fall below the anchored window's $0.964$ ($0.953$ unbounded, $0.961$ proposed), so a cadence coarser than $\sim\!50$ visits forfeits the gain of retained memory.pdf

Two robustness checks support attributing the gain to the adaptation policy. First, extending the passive window to $L_w\!=\!2000$ raises its best coverage to $0.979$, below the proposed method's $0.984$ despite $6\%$ more channel uses. Second, replacing the NP-CDM with a false-alarm-matched, block-wise Page--Hinkley detector reduces anchored-window coverage by $1.3$ and $1.7$ percentage points on D1 and D3, with its boundary resolved only to a $100$-sample block, but leaves the proposed method's coverage within the CI and resource use within $1\%$, at $12$ against $7$ epochs per run. Detector choice therefore affects where a discarding policy restarts, whereas the gain comes from event-triggered state-space re-derivation and re-binning with bounded memory.

Two qualifications remain. First, the extreme bin edges in \eqref{eq:adaptomega} rest on few post-change samples, inherent to any empirical-quantile rule at $L\!=\!15$, which costs on the return leg of D8 (Table~\ref{tab:d8}). Second, $\eta\!=\!0.99$ is a comparative benchmark, not an end-to-end $10^{-5}$ certificate: verifying $\epsilon$ directly needs $\sim\!\!10^{7}$ post-drift samples per method and scenario.

\section{Conclusion}\label{sec:concl}
Unannounced drift makes a frozen EVT/KDE predictor under-cover silently and makes cumulative refitting on a fixed state space resource-inefficient. The two losses have different sources and remedies: a stale partition forfeits conditional resolution, which re-deriving it at detected changes restores, whereas discarding history forfeits tail support, which retaining observations under the new partition preserves. The design principle that follows is to adapt the state representation without discarding the tail observations on which extreme-quantile accuracy depends, and the proposed method does both within a bounded per-state memory. At matched achieved coverage, it saves $4$--$18\%$ of channel uses on the five representation-changing scenarios, changes
stationary and load-only resource use by at most $0.4\%$, and needs $O(KL)$ rather than $O(t)$ storage. Recognizing returning regimes, finite-sample coverage bounds, and validation on measured interference traces remain as possible future extensions.

\appendices
\section{Proof of Proposition~\ref{prop:cap}}
\label{sec:analysis}
\vspace{-3pt}
\begin{proof}
Fix a state $r$ of $\Omega^{(k)}$. Since the $b_r^{(k)}$ inherited samples are older than every sample appended
after the epoch update, FIFO eviction removes them first, so after $v_r^{(k)}(t)$ visits at most $[K-v_r^{(k)}(t)]_{+}$ inherited and at most $s_r^{(k)}$ stale samples remain among the $m_r^{(k)}(t)$ held, which gives \eqref{eq:alpha_cap} and $\alpha_r^{(k)}(t)=0$ once $v_r^{(k)}(t)\ge K$. The bound is on the buffer the update produces: re-binning does not recover samples discarded under the previous partition, so $b_r^{(k)}$ may fall short of a prefix re-bin and the variance term through $m_r^{(k)}(t)$ is conservative. The remainder derives \eqref{eq:mse_cap}, which locally replaces the estimation term in \eqref{eq:decomp} with a bias and a variance term.

\smallskip\noindent\textbf{Bias.} Under linear-contamination approximation, the pseudo-true fitted quantile of a mixed sample interpolates between regimes, so $\mathrm{Bias}_r(t)\approx\alpha_r^{(k)}(t)\,\Delta Q_{\eta,r}$. The approximation is first order, since a mixture of two GPDs is
not a GPD and the single-GPD fit can be misspecified at intermediate contamination; the LMRD diagnostic of
Sec.~\ref{ssec:lmrd} checks its adequacy in operation.
 
\smallskip\noindent\textbf{Variance.} Under the PWM asymptotic regime \cite{hosking1987pwm}, propagating the parameter variances through \eqref{eq:evt} gives, for a fit on $N_u$ exceedances,
\begin{equation}
\mathrm{Var}\big(\hat q_{\eta}\big)\approx
\frac{J_{\eta}^{\top}\Sigma_{\theta}J_{\eta}}{N_u}
=\frac{\kappa_{\eta}(\xi,\rho_u)\,\sigma_{u}^{2}}{N_u}, \,
\kappa_{\eta}:=\frac{J_{\eta}^{\top}\Sigma_{\theta}J_{\eta}}{\sigma_u^{2}},
\label{eq:varQ}
\end{equation}
with $\Sigma_{\theta}$ the PWM asymptotic covariance \cite[Eq.~(16)]{hosking1987pwm}, $J_{\eta}$ the quantile Jacobian with respect to $(\sigma_u,\xi)$, and $\kappa_{\eta}$ bounded on $\xi\in[-1/2,1/2)$. The buffer supplies $N_u=\rho_{u_r}\, m_r^{(k)}(t)$ exceedances in expectation. Fading correlation and message duration make consecutive observations dependent, so \eqref{eq:varQ} treats $N_u$ as if independent, a first-order approximation that understates the variance by a factor set by the correlation length. Adding squared bias and variance yields \eqref{eq:mse_cap}. At the evaluated operating point, $\kappa_{\eta}(\xi,\rho_u)$ lies
in $[2,5]$ over $\xi\in[0,0.25]$.
\end{proof}

\section{The Change-Point-Anchored Accumulating Window}
\label{sec:S-window}
This appendix specifies the comparison policy that
discards pre-change history at each accepted detection, the
\emph{anchored window} of Sec.~\ref{ssec:memory} and Table~\ref{tab:main}. It is constructed as the strongest form of discarding, so
that the finding of Sec.~\ref{sec:results}, that retaining and re-binning
history is better, is measured against a discarding policy given
every advantage. The detector, the refractory rule and the epoch
definition, \eqref{eq:epoch}, are unchanged.

Given the anchor $\hat t^{\ast}$ of an accepted detection at
$\tau_d$, a discarding policy must choose its retraining window.
Two extremes bracket the design. The \emph{passive} window
$\mathcal{W}_{\mathrm{P}}(t)=[t\!-\!L_w\!+\!1,\,t]$, a fixed length
slid forward each slot, ignores $\hat t^{\ast}$ and suffers the.pdf
three consequences of Sec.~\ref{sec:failure}-D. A
\emph{hard-trim} window
$\mathcal{W}_{\mathrm{HT}}=[\hat t^{\ast}\!+\!1,\,\tau_d]$, frozen
at the moment of detection, contains no sample before the
estimated boundary but holds too few. State $r$ receives only about
$(\tau_d-\hat t^{\ast})\,\pi_r\,\rho_{u_r}$ exceedances, with
$\pi_r$ its occupancy and $\rho_{u_r}$ its exceedance rate
(Sec.~\ref{ssec:mixture}), and freezing the window
at $\tau_d$ forgoes the post-change samples that arrive later.
On D1, the hard-trim policy is worse than passive (Sec.~\ref{sec:S-crossover}).

The anchored window combines the two strengths. Within epoch $k$,
for $\tau^{(k)}\!\le\!t\!<\!\tau^{(k+1)}$, it is
\begin{equation}
\mathcal{W}_k(t)=\big\{\,r:\ \max\!\big(a^{(k)}\!+\!1,\,
t-L_{\mathrm{win}}\!+\!1\big)\le r\le t\,\big\},
\label{eq:S-awin}
\end{equation}
so that $|\mathcal{W}_k(t)|=\min\{t-a^{(k)},L_{\mathrm{win}}\}$. It
is pinned at the anchor, so no sample before the estimated boundary
is re-ingested, although an early anchor may admit $O(W)$
pre-change samples. It grows with $t$ until it reaches
$L_{\mathrm{win}}$ and then slides, which bounds its exposure to
slow drift within an epoch. The anchor slot $a^{(k)}$ itself, the
end of the last all-clear block and hence a pre-change sample, is
excluded. The mixture model, the per-state
$(u_i,\sigma_{u_i},\xi_i,\rho_{u_i})$ of \eqref{eq:evt} together with the bulk KDE, is refit on $\mathcal{W}_k(t)$
whenever
\begin{equation}
(t-t_{\text{last refit}})\ge P_{\text{refit}}
\ \ \text{and}\ \ |\mathcal{W}_k(t)|\ge n_{\min},
\label{eq:S-refit}
\end{equation}
with $P_{\text{refit}}\!=\!200$ slots, $n_{\min}\!=\!200$ and
$L_{\mathrm{win}}\!=\!2000$ (Table~\ref{tab:params}), where the second condition defers the first refit until the
window reaches its minimum operational size, and the latest
estimates are held between refits. States with too few exceedances for a tail fit
fall back to the empirical rule of Sec.~\ref{ssec:lmrd}. Without an accepted alarm the initial fit persists, so
stationarity carries no penalty and refits occur only at the
measured false-alarm rate (Sec.~\ref{ssec:discussion}). After each accepted detection, the reference chain is
re-learned on post-anchor data and the search restarts. At every
accepted event, the Table~\ref{tab:main} anchored baseline re-derives
$\Omega$ and selects $\hat q$ with the same epoch-update and tail
machinery as the proposed method; only its predictor memory is
replaced by $\mathcal{W}_k(t)$. The fixed-$\Omega$ variant used in
Proposition~\ref{prop:S1} and Fig.~\ref{fig:S-mse} is solely a
diagnostic that isolates the window operator.

\section{First-Order Crossover Against Passive Refitting}
\label{sec:S-crossover}
The anchored window removes contamination but starts small, so
early in an epoch its tail estimate is variance-limited while the
passive estimate is bias-limited. From which sample count onward is
the anchored estimate the better one? Comparing the first-order
predicted-quantile MSE of the two windows for one state gives a
sufficient condition on the anchored sample count, stated and
proved next and examined empirically in Fig.~\ref{fig:S-mse}.
Notation follows Sec.~\ref{ssec:memory}: subscripts
$0$ and $1$ denote pre- and post-change GPD parameters,
$\Delta Q_{\eta,r}$ the resulting shift of the state's
$\eta$-quantile, and $\kappa_{\eta}(\xi)$ the dimensionless
variance factor of \eqref{eq:varQ}, Appendix~\ref{sec:analysis}'s
$\kappa_{\eta}(\xi,\rho_{u_r})$ with its second argument
suppressed. In addition,
$\alpha_{\mathrm{P}}(t)$ is the pre-change fraction of the passive
window, the analogue for a window of the buffer stale fraction
$\alpha_r^{(k)}(t)$ of Sec.~\ref{ssec:memory}.

\begin{prop}[First-order crossover]
\label{prop:S1}
Fix $t\!>\!t^{\ast}$ and one state $r$. Let
$n(t)\!=\!\min\{t-\hat t^{\ast},\,L_{\mathrm{win}}\}$ be the
anchored window's slot count and $\lambda_{u}\!=\!\pi_{r}\rho_{u_r}$
the state's per-slot exceedance rate, taken common to both regimes
as a simplifying approximation. Assume the regularity conditions
of~\cite{hosking1987pwm}, a localization error
$|\hat t^{\ast}\!-\!t^{\ast}|\!=\!O(W)$, and the linear-contamination
approximation for the passive window's bias; approximate the
passive window's variance coefficient by its pre-change value
$\kappa_{\eta}(\xi_0)\sigma_0^2$ during the crossover interval
(once the window is fully post-change the same comparison holds
with the post-change coefficient); and neglect the
$O\big((W/n(t))^{2}|\Delta Q_{\eta,r}|^{2}\big)$ contribution of an
early anchor. Let $\mathrm{MSE}_{\mathrm{P}}(t)$ and
$\mathrm{MSE}_{\mathrm{A}}(t)$ denote the first-order MSE of the
predicted quantile for state $r$ under the passive and anchored
windows. Then, under this variance proxy,
$\mathrm{MSE}_{\mathrm{A}}(t)\!\le\!\mathrm{MSE}_{\mathrm{P}}(t)$
whenever
\begin{equation}
n(t)\;\ge\;n_{0}(t)\;:=\;
\frac{\kappa_{\eta}(\xi_{1})\,\sigma_{1}^{2}}
{\lambda_{u}\!\left[\alpha_{\mathrm{P}}(t)^{2}|\Delta Q_{\eta,r}|^{2}\right]
+\dfrac{\kappa_{\eta}(\xi_{0})\,\sigma_{0}^{2}}{L_{w}}}.
\label{eq:S-n0}
\end{equation}
In the early transient after detection ($\alpha_{\mathrm{P}}(t)\!\simeq\!1$),
dropping the passive-variance term simplifies \eqref{eq:S-n0}
conservatively to
\begin{equation}
n_{0}\;\approx\;\frac{\kappa_{\eta}(\xi_{1})\,\sigma_{1}^{2}}
{\lambda_{u}\,|\Delta Q_{\eta,r}|^{2}}.
\label{eq:S-n0worst}
\end{equation}
\end{prop}

The proposition is conditional on an accepted detection and an
$O(W)$ localization error. No detection-delay or false-alarm
guarantee is claimed. The $O(W)$ assumption describes the typical
run: on the abrupt D1 and D7 the median absolute anchor error is $7$--$9$
blocks (Sec.~\ref{ssec:discussion}), while in
roughly one run in ten the change falls inside the refractory of a
preceding alarm and the first accepted epoch belongs to a later
detection. The proposition does not cover those runs.

\begin{proof}
Fix one state and compare the passive window
$\mathcal{W}_{\mathrm{P}}(t)$ with the anchored window
\eqref{eq:S-awin} for $t\!>\!t^{\ast}$, where the true parameters
have shifted from $(\sigma_{0},\xi_{0})$ to $(\sigma_{1},\xi_{1})$.
A window of $|\mathcal{W}|$ slots holds $|\mathcal{W}|\lambda_{u}$
exceedances in expectation, and
$\Delta Q_{\eta,r}=Q_{\eta,1}-Q_{\eta,0}$ is the state's
$\eta$-quantile, \eqref{eq:evt}, under the
post-change minus the pre-change parameters.

\smallskip\noindent\textbf{Bias.} For
$t\in[t^{\ast},t^{\ast}\!+\!L_{w}]$ the passive window carries the
pre-change fraction
$\alpha_{\mathrm{P}}(t)\!=\!\big(1-(t-t^{\ast})/L_{w}\big)_{+}$, zero
thereafter. Under the linear-contamination approximation the fitted
quantile of the mixed window interpolates linearly between the
regimes,
\begin{equation}
\mathrm{Bias}_{\mathrm{P}}(t)\;\approx\;\alpha_{\mathrm{P}}(t)\,\Delta Q_{\eta,r}.
\label{eq:biasP}
\end{equation}
This is an approximation rather than a consequence of PWM
regularity, since a mixture of two GPDs is generally not a GPD
(see Appendix~\ref{sec:analysis}), and $\alpha_{\mathrm{P}}(t)$ is the
slot-level fraction, which equals the exceedance fraction only when
the two regimes have equal exceedance rates. The anchored
window contains only post-change exceedances whenever
$\hat t^{\ast}\!\ge\!t^{\ast}$. Under the localization assumption an
early anchor admits a residual fraction
$(t^{\ast}\!-\!\hat t^{\ast})_{+}/n(t)$ and a bias of that order
times $\Delta Q_{\eta,r}$, vanishing as the window grows.

\smallskip\noindent\textbf{Variance.} In the PWM asymptotic
regime~\cite{hosking1987pwm}, propagating the parameter variances
through \eqref{eq:evt} gives, for a window with
exceedance count $N_{u}$,
\begin{equation}
\mathrm{Var}\big(\hat q_{\eta}\big)
\approx \frac{\kappa_{\eta}(\xi)\,\sigma^{2}}{N_{u}},
\qquad
\kappa_{\eta}=J^{\top}\Sigma_{\!\theta}J/\sigma^{2},
\label{eq:varQ-S}
\end{equation}
where $\Sigma_{\!\theta}$ is the PWM asymptotic
covariance~\cite[Eq.~(16)]{hosking1987pwm}, $J$ the Jacobian of the
quantile with respect to $(\sigma,\xi)$, and $\kappa_{\eta}(\xi)$
bounded on $\xi\!\in\![-1/2,1/2)$. Exceedances are treated as
independent or sufficiently mixing, with $N_{u}$ read as an
effective count (Sec.~\ref{ssec:mixture}). The
Monte-Carlo evaluation itself retains the full temporal dependence.
At the operating point ($\rho_{u}\!=\!0.03$, $1\!-\!\eta\!=\!0.01$),
$\kappa_{\eta}(\xi)$ lies in $[2,5]$ over $\xi\in[0,0.25]$, the
range used in Fig.~\ref{fig:S-mse}.

\smallskip\noindent\textbf{Crossover.} Adding squared bias and
variance, with $N_{u}\!=\!L_{w}\lambda_{u}$ for passive and
$N_{u}\!=\!n(t)\lambda_{u}$ for anchored,
\begin{align}
\mathrm{MSE}_{\mathrm{P}}(t) &= \alpha_{\mathrm{P}}(t)^{2}\,|\Delta Q_{\eta,r}|^{2}
+ \frac{\kappa_{\eta}(\xi_{0})\,\sigma_{0}^{2}}{L_{w}\lambda_{u}},
\label{eq:mseP}\\
\mathrm{MSE}_{\mathrm{A}}(t) &= \frac{\kappa_{\eta}(\xi_{1})\,\sigma_{1}^{2}}
{n(t)\,\lambda_{u}}.
\label{eq:mseA}
\end{align}
The inequality $\mathrm{MSE}_{\mathrm{A}}\!\le\!\mathrm{MSE}_{\mathrm{P}}$
rearranges to $n(t)\!\ge\!n_{0}(t)$ with $n_{0}(t)$ as in
\eqref{eq:S-n0}. For $\alpha_{\mathrm{P}}(t)\!\simeq\!1$, dropping the
positive passive-variance term gives the conservative approximation
\eqref{eq:S-n0worst}, which is not a uniform bound over all
$\alpha_{\mathrm{P}}$.
\end{proof}

\subsection{Empirical test of the crossover}
\begin{figure}[t]
\centering
\includegraphics[width=0.85\columnwidth]{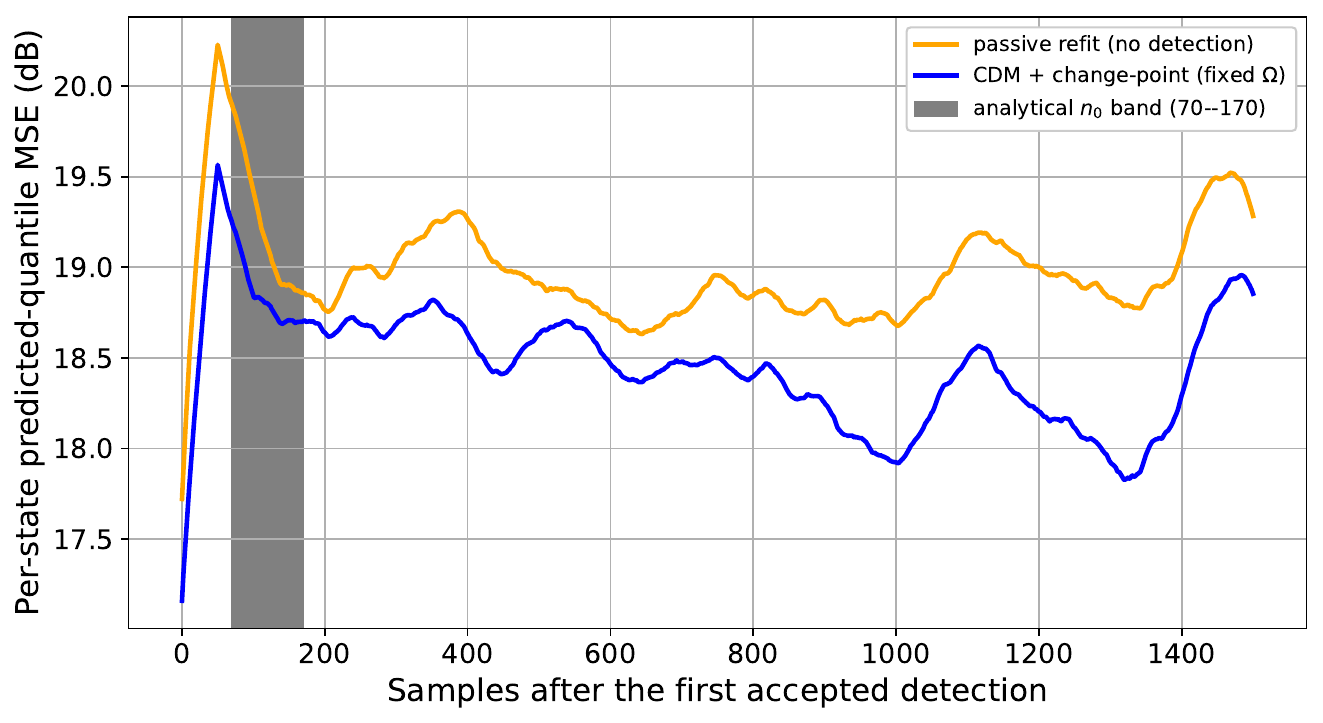}
\vspace{-8pt}
\caption{D1 post-detection per-state quantile MSE (computed in
linear INR units, displayed in dB) for passive refitting and the
fixed-$\Omega$ anchored window, aligned on the first accepted
detection. The shaded interval shows the width of Proposition~\ref{prop:S1}'s crossover range $n_{0}\!\in\![70,170]$; it is counted from the anchor, so its true position lies about $n(\tau_d)$ samples to the left, before the origin in most runs.}
\label{fig:S-mse}
\end{figure}

Fig.~\ref{fig:S-mse} tests the crossover of
Proposition~\ref{prop:S1}: from which sample count onward
decontamination outweighs the anchored window's smaller sample. The
diagnostic is evaluated at the inherited $97$th-percentile
threshold, isolating the window operator from threshold selection.
At that operating point, the proposition predicts dominance once the
anchored window holds
$n(t)\!=\!t-\hat t^{\ast}\!\ge\!n_{0}\!\approx\!70$--$170$ samples.
At detection, the anchored window holds $n(\tau_d)=\tau_d-\hat t^{\ast}$ samples, which equals the detection delay only for an exact anchor. On D1, the median is $270$; every run has at least $70$ samples, and $81\%$ have at least $170$. Thus, the approximate crossover is reached by detection in most runs, though it may occur later in others. Consistent with this prediction, the fixed-$\Omega$ anchored window has lower measured MSE than passive refitting throughout the plotted horizon, although the gap is not monotone. The anchored window with adaptive
$\Omega$ is omitted from this diagnostic, since its per-epoch
mapping does not share the fixed-$\Omega$ reference quantile.

Replacing the accumulating anchor with the frozen hard-trim window
of Sec.~\ref{sec:S-window} yields, on D1, coverage $0.873$, worse
than passive's $0.930$ and with larger variance. Hard-trim
decontaminates but does not accumulate, and passive accumulates but
does not decontaminate. The anchored window does both, which is
what makes it the discarding policy to compare against.

\section{Detection Delay by Scenario}
\label{sec:S-delay}
\begin{figure}[t]
\centering
\includegraphics[width=1\columnwidth]{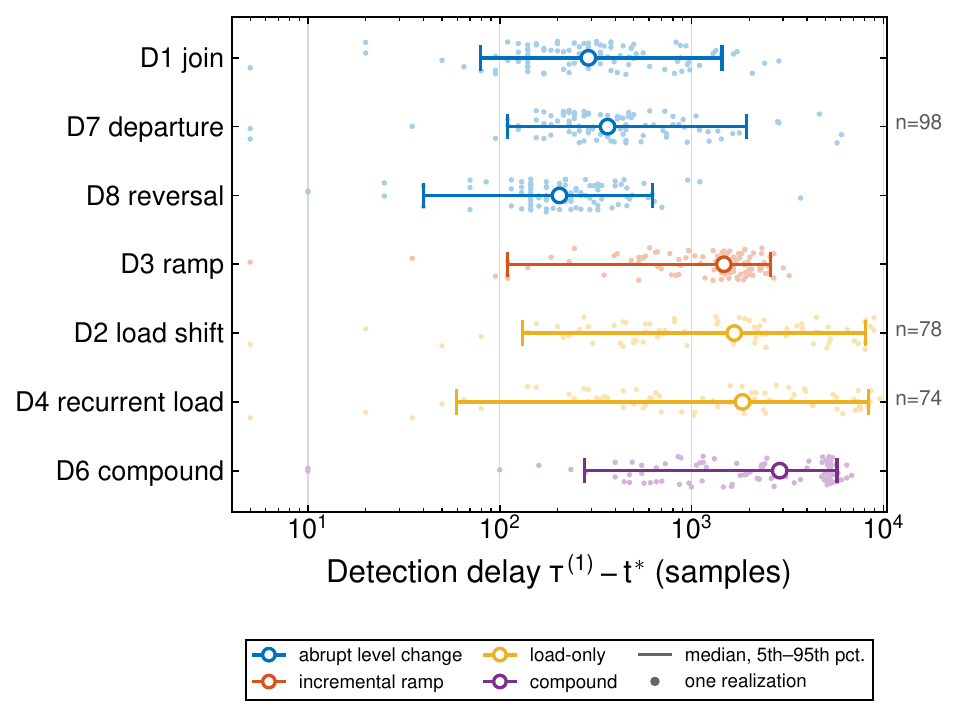}
\vspace{-8pt}
\caption{Detection delay of the first accepted detection after
$t^{\ast}$, by scenario: median (circle), 5th--95th percentile
(bar) and individual realizations (dots), $100$ realizations on
seeds disjoint from the evaluation. D2, D4 and D7 show only the runs
that detected; the detector misses the change in $22\%$, $26\%$ and
$2\%$ of runs respectively. The compound D6 is bimodal: about half
the runs detect the mild load shift, the rest first fire at the
join $5\!\times\!10^{3}$ samples later.}
\label{fig:S-recov}
\end{figure}

Fig.~\ref{fig:S-recov} gives the detection delay, the samples from
onset to the first accepted detection, by scenario. Delay is a
property of the detector and is shared by every detection-driven
method of Sec.~\ref{sec:results}, and it tracks how sharply a scenario
moves the state occupancy rather than how large the change is. An
abrupt step in interference level is detected within a few hundred
samples, whereas a gradual ramp or a load-only shift, which moves
occupancy slowly, takes an order of magnitude longer and, for the
load-only D2 and D4, is missed altogether in about a quarter of the
runs. On the compound D6, the mild load shift is caught in about
half the runs. In the rest, the first accepted detection is the
interferer join $5\!\times\!10^{3}$ samples later, which makes the
delay distribution bimodal.

\section{Window Lengths $L_w$ and $L_{\mathrm{det}}$}
\label{sec:S-windows}
\begin{figure}[t]
\centering
\subfloat[]{\includegraphics[width=0.9\columnwidth]{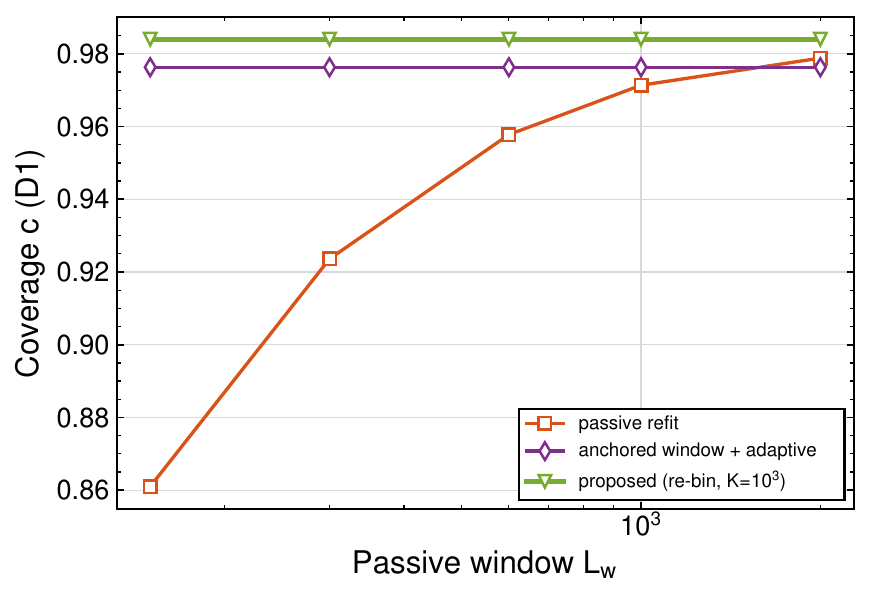}}\hfil
\subfloat[]{\includegraphics[width=0.9\columnwidth]{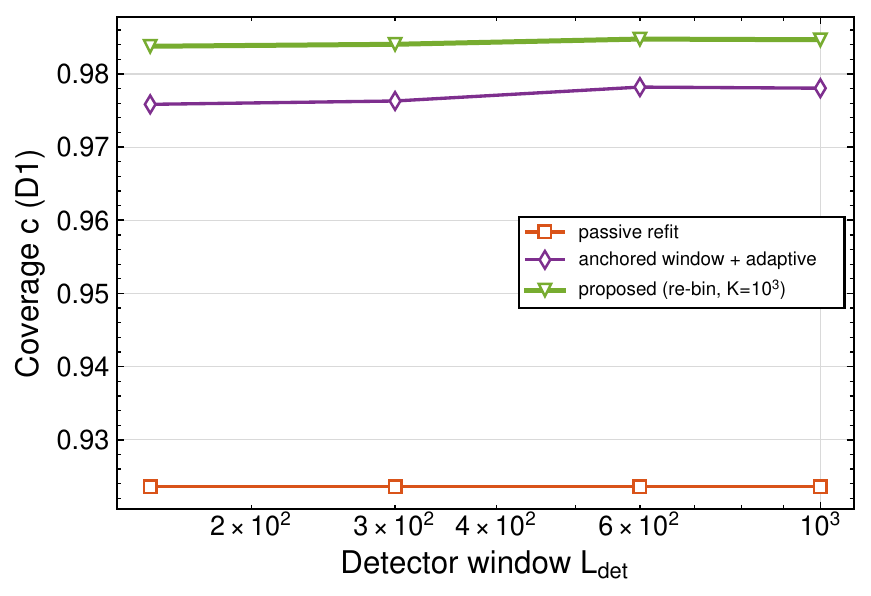}}
\vspace{-6pt}
\caption{Window sensitivity on D1 ($20$ realizations):
(a)~passive window $L_w$ with the detector pinned at
$L_{\mathrm{det}}\!=\!300$, so only passive moves; (b)~detector
window $L_{\mathrm{det}}$ with passive pinned.}
\label{fig:windows}
\vspace{-6pt}
\end{figure}

Two window lengths enter the comparison: the passive baseline's
refit window $L_w$ and the detector's trailing window
$L_{\mathrm{det}}$ (Table~\ref{tab:params}). Each is
swept on D1 with the other held at its default, so that only one
method moves at a time (Fig.~\ref{fig:windows}).

\textit{Passive refit window.} Over
$L_{w}\!\in\!\{150,300,600,1000,2000\}$, passive's coverage rises
monotonically, to $0.979$ at $L_w\!=\!2000$ on D1, above the
anchored window's $0.976$ but below the proposed method's $0.984$,
which the proposed method attains at $6\%$ fewer channel uses. Long
windows reach the highest settled coverage on every swept scenario
($0.970$--$0.981$ at $L_{w}\!=\!2000$ on D1, D3 and D5), because on
upward drift the exceedances self-select the post-change samples
into the tail fit, so even a contaminated long window adapts its
tail quickly. They pay for this in resource cost under
representation-changing drift ($1.43$ against $1.23$ for the
adaptive partition on D3) and in a window length that must be tuned
against unknown drift timescales and directions, which the anchored
window avoids by being short immediately after a change and long
thereafter. A longer window does not close the gap, so the passive
baseline of Sec.~\ref{sec:results} is not handicapped by its default.

\textit{Detector window.} Over $L_{\mathrm{det}}\!\in\![150,1000]$,
the anchored window's coverage moves from $0.976$ to $0.978$ and
the proposed method's by $0.001$, the same insensitivity to the
detector's parameters that the $(h,\rho)$ sweep of
Sec.~\ref{ssec:discussion} shows.

\section{Page--Hinkley Detector Baseline}
\label{sec:S-ph}
Sec.~\ref{ssec:discussion} replaces the NP-CDM with a
Page--Hinkley baseline to test whether the gains depend on the
particular detector. The baseline runs on standardized means of
nonoverlapping $100$-sample blocks. On that block sequence it runs
Page's two-sided recursion,
\begin{equation}
g^{\pm}_{b}=\max\big(0,\;g^{\pm}_{b-1}\pm z_{b}-\delta\big),
\label{eq:S-ph}
\end{equation}
and alarms when either statistic exceeds $\lambda$. We set
$\delta\!=\!0.3$ and $\lambda\!=\!10$ on $20$ stationary traces
disjoint from the evaluation, so that the baseline raises no more
false alarms than the proposed detector. The minimizing block gives
the boundary, and everything downstream is unchanged. Blocking is
necessary, since applied per sample the recursion over-fires by an
order of magnitude on this correlated stream.

Driving the fixed-$\Omega$ anchored window, the baseline detects
later on D1 (mean delay $572$ against the NP-CDM's $462$ samples)
and loses $1.3$ and $1.7$ percentage points of coverage on D1 and
D3, with separated CIs. The extra delay is a negligible share of
the post-change horizon; the cause is boundary resolution: the minimizing block locates the
change only to its own $100$ samples, against the $W\!=\!15$
samples of the NP-CDM's subsequence, so the anchored window opens
too early and retains pre-change samples that its discard policy
would otherwise have removed. Driving the proposed policy, in which history is retained and
re-binned, the same boundary error is immaterial: the baseline
matches the NP-CDM on coverage within the CI ($0.986$ against
$0.985$ on D1, $0.973$ on both on D3), at resource use within $1\%$,
and with $3$--$12\%$ lower pinball loss, coinciding with the more frequent
re-derivation its $12$ accepted epochs per run produce against the
NP-CDM's $7$. Boundary resolution is therefore much less
consequential when history is retained and re-binned: in the
evaluated scenarios the false-alarm-matched Page--Hinkley detector
gives statistically indistinguishable coverage and resource use
within $1\%$, so the principal gain comes from the downstream
adaptation policy rather than the particular detector. The NP-CDM
is retained because it operates on the state
stream the predictor already maintains
(Sec.~\ref{ssec:discussion}).

\bibliographystyle{IEEEtran}
\bibliography{refs}

@IEEEtranBSTCTL{IEEEexample:BSTcontrol,
  CTLuse_forced_etal       = "yes",
 CTLmax_names_forced_etal = "2",
  CTLnames_show_etal       = "1" 
}

@inproceedings{salehi2023reliable,
  author    = {Salehi, Fateme and Mahmood, Aamir and Mahmood, Nurul Huda and
               Gidlund, Mikael},
  title     = {Reliable Interference Prediction and Management with
               Time-Correlated Traffic for {URLLC}},
  booktitle = {IEEE GLOBECOM},
  year      = {2023}
}

@INPROCEEDINGS{salehi2025evt,
  author={Salehi, Fateme and Mahmood, Aamir and Coleri, Sinem and Gidlund, Mikael},
  booktitle={IEEE ICC}, 
  title={Ultra-High Reliability by Predictive Interference Management Using Extreme Value Theory}, 
  year={2025},
  volume={},
  number={},
  pages={2538-2543},
  doi={10.1109/ICC52391.2025.11161826}
}

@article{mahmood2020predictive,
  author  = {Mahmood, Nurul Huda and L\'{o}pez, Onel A. and Alves, Hirley
             and Latva-aho, Matti},
  title   = {A Predictive Interference Management Algorithm for {URLLC} in
             Beyond 5{G} Networks},
  journal = {IEEE Commun. Lett.},
  volume  = {25},
  number  = {3},
  pages   = {995--999},
  year    = {2020}
}

@ARTICLE{mehrnia2024nonstat,
  author={Mehrnia, Niloofar and Coleri, Sinem},
  journal={IEEE Trans. Veh. Technol.}, 
  title={Non-Stationary Wireless Channel Modeling Approach Based on Extreme Value Theory for Ultra-Reliable Communications}, 
  year={2021},
  volume={70},
  number={8},
  pages={8264-8268},
  doi={10.1109/TVT.2021.3091378}
}

@ARTICLE{roveri2019dtmc,
  author={Roveri, Manuel},
  journal={IEEE Trans. Neural Netw. Learn. Syst.}, 
  title={Learning Discrete-Time {M}arkov Chains Under Concept Drift}, 
  year={2019},
  pages={2570-2582},
  doi={10.1109/TNNLS.2018.2886956}
}

@article{polyanskiy2010fbl,
  author  = {Polyanskiy, Yury and Poor, H. Vincent and Verd\'{u}, Sergio},
  title   = {Channel Coding Rate in the Finite Blocklength Regime},
  journal = {IEEE Trans. Inf. Theory},
  volume  = {56},
  number  = {5},
  pages   = {2307--2359},
  year    = {2010}
}

@article{bennis2018tail,
  author  = {Bennis, Mehdi and Debbah, M\'{e}rouane and Poor, H. Vincent},
  title   = {Ultrareliable and Low-Latency Wireless Communication: Tail,
             Risk, and Scale},
  journal = {Proc. IEEE},
  volume  = {106},
  number  = {10},
  pages   = {1834--1853},
  year    = {2018}
}

@book{coles2001evt,
  author    = {Coles, Stuart},
  title     = {An Introduction to Statistical Modeling of Extreme Values},
  publisher = {Springer},
  year      = {2001}
}

@article{page1954cusum,
  author  = {Page, E. S.},
  title   = {Continuous Inspection Schemes},
  journal = {Biometrika},
  volume  = {41},
  number  = {1/2},
  pages   = {100--115},
  year    = {1954}
}

@article{hosking1987pwm,
  author  = {Hosking, J. R. M. and Wallis, J. R.},
  title   = {Parameter and Quantile Estimation for the Generalized {Pareto}
             Distribution},
  journal = {Technometrics},
  volume  = {29},
  number  = {3},
  pages   = {339--349},
  year    = {1987}
}

@article{hosking1990lmoments,
  author  = {Hosking, J. R. M.},
  title   = {{L}-Moments: Analysis and Estimation of Distributions Using
             Linear Combinations of Order Statistics},
  journal = {J. R. Stat. Soc., B: Stat. Methodol.},
  volume  = {52},
  number  = {1},
  pages   = {105--124},
  year    = {1990}
}

@article{gama2014survey,
  author  = {Gama, Jo{\~a}o and \v{Z}liobait\.{e}, Indr\.{e} and
             Bifet, Albert and Pechenizkiy, Mykola and
             Bouchachia, Abdelhamid},
  title   = {A Survey on Concept Drift Adaptation},
  journal = {ACM Computing Surveys},
  volume  = {46},
  number  = {4},
  pages   = {Article 44, 1--37},
  year    = {2014}
}

@inproceedings{bifet2007adwin,
  author    = {Bifet, Albert and Gavald\`{a}, Ricard},
  title     = {Learning from Time-Changing Data with Adaptive Windowing},
  booktitle = {Proc. SIAM International Conference on Data Mining (SDM)},
  pages     = {443--448},
  year      = {2007}
}

@inproceedings{gama2004ddm,
  author    = {Gama, Jo{\~a}o and Medas, Pedro and Castillo, Gladys and
               Rodrigues, Pedro},
  title     = {Learning with Drift Detection},
  booktitle = {Advances in Artificial Intelligence -- SBIA 2004},
  series    = {Lecture Notes in Computer Science},
  pages     = {286--295},
  publisher = {Springer},
  year      = {2004}
}

@misc{adams2007bocpd,
  author = {Adams, Ryan Prescott and MacKay, David J. C.},
  title  = {Bayesian Online Changepoint Detection},
  year   = {2007},
  note   = {arXiv:0710.3742}
}

@book{basseville1993detection,
  author    = {Basseville, Mich\`{e}le and Nikiforov, Igor V.},
  title     = {Detection of Abrupt Changes: Theory and Application},
  publisher = {Prentice Hall},
  year      = {1993}
}

@article{lorden1971procedures,
  author  = {Lorden, Gary},
  title   = {Procedures for Reacting to a Change in Distribution},
  journal = {The Annals of Mathematical Statistics},
  volume  = {42},
  number  = {6},
  pages   = {1897--1908},
  year    = {1971}
}

@article{landwehr1979pwm,
  author  = {J. M. Landwehr and N. C. Matalas and J. R. Wallis},
  title   = {Probability weighted moments compared with some traditional
             techniques in estimating {G}umbel parameters and quantiles},
  journal = {Water Resources Research},
  volume  = {15},
  number  = {5},
  pages   = {1055--1064},
  year    = {1979}
}

@article{botev2010kde,
  author  = {Z. I. Botev and J. F. Grotowski and D. P. Kroese},
  title   = {Kernel density estimation via diffusion},
  journal = {The Annals of Statistics},
  volume  = {38},
  number  = {5},
  pages   = {2916--2957},
  year    = {2010}
}

@ARTICLE{gautam2025evt,
  author={Gautam, Pramesh and Sapkota, Sushmita and Bockelmann, Carsten and Raj Pandey, Shashi and Dekorsy, Armin},
  journal={IEEE Open J. Commun. Soc.}, 
  title={Extreme Value Theory-Based Distributed Interference Prediction for {6G} Industrial Sub-Networks}, 
  year={2025},
  volume={6},
  number={},
  pages={9121-9143},
  doi={10.1109/OJCOMS.2025.3625505}
}

@article{malik2020nbiot,
  author  = {Malik, Hassan and Alam, Muhammad Mahtab and Pervaiz, Haris and Le Moullec, Yannick and Al-Dulaimi, Anwer and Parand, Sven and Reggiani, Luca},
  journal = {IEEE Netw.},
  title   = {Radio Resource Management in {NB-IoT} Systems: Empowered by Interference Prediction and Flexible Duplexing},
  year    = {2020}, volume  = {34}, number  = {1}, pages   = {144--151}
}

@article{schmidt2021interference,
  author  = {Schmidt, Jorge F. and Schilcher, Udo and Atiq, Mahin K. and Bettstetter, Christian},
  journal = {IEEE Trans. Veh. Technol.},
  title   = {Interference Prediction in Wireless Networks: Stochastic Geometry Meets Recursive Filtering},
  year    = {2021}, volume  = {70}, number  = {3}, pages   = {2783--2793}
}

@article{kim2019sense,
  author  = {Kim, Sunghyun and Cha, Hyeryung and Kim, Jihong and Ko, Seung-Woo and Kim, Seong-Lyun},
  journal = {IEEE Trans. Wirel. Commun.},
  title   = {Sense-and-Predict: Harnessing Spatial Interference Correlation for Cognitive Radio Networks},
  year    = {2019}, volume  = {18}, number  = {5}, pages   = {2777--2793}
}

@article{brighente2022interference,
  author  = {Brighente, Alessandro and Mohammadi, Jafar and Baracca, Paolo and Mandelli, Silvio and Tomasin, Stefano},
  journal = {IEEE Trans. Wirel. Commun.},
  title   = {Interference Prediction for Low-Complexity Link Adaptation in Beyond {5G} Ultra-Reliable Low-Latency Communications},
  year    = {2022}, volume  = {21}, number  = {10}, pages   = {8403--8415}
}

@article{koenker1978,
  author  = {Koenker, Roger and Bassett, Gilbert},
  title   = {Regression quantiles},
  journal = {Econometrica},
  volume  = {46}, number = {1}, pages = {33--50}, year = {1978}}

\end{document}